\newif\ifpreprint
 
\preprinttrue 
  
\ifpreprint
\documentclass[journal=jctcce,manuscript=article]{achemso}
\else
\documentclass[journal=jctcce,manuscript=article,layout=twocolumn]{achemso}
\fi

\usepackage[colorlinks = true,
            linkcolor = blue,
            urlcolor  = black,
            citecolor = blue,
            anchorcolor = black]{hyperref}

\usepackage{amsmath}
\usepackage{newtxtext,newtxmath}
\usepackage{graphicx}
\usepackage{bm}
\usepackage{braket}
\usepackage{threeparttable}
\usepackage{multirow}
\usepackage[format=plain,singlelinecheck=false,font=footnotesize,labelfont=bf,labelsep=space]{caption}
\usepackage{dcolumn}
\usepackage{float}
\usepackage{xspace}
\usepackage{ifthen}
\usepackage{dsfont}
\usepackage{algorithm,algpseudocode}
\usepackage{longtable}
\usepackage{color,soul}
\usepackage[usenames,dvipsnames]{xcolor}
\usepackage{fancyvrb}
\usepackage{adjustbox}
\usepackage{setspace}
\usepackage{indent first}

\usepackage{changepage}

\usepackage[version=4]{mhchem} 
\usepackage[T1]{fontenc}       

\usepackage{titlesec}
\usepackage{footnote}

\usepackage[fontsize=11pt]{scrextend}
\titleformat{\section}
  {\normalfont\sffamily\bfseries\color{OliveGreen}}
  {\thesection.}{0.25em}{\uppercase}
  
  \titleformat{\subsection}
  {\normalfont\sffamily\bfseries}
  {\thesubsection}{0.25em}{}
  
  \titleformat{\suppinfo}
  {\normalfont\sffamily\bfseries}
  {\thesubsection}{0.25em}{}

\newcommand*{\Eh}{$E_{\rm h}$\xspace}
\newcommand*{\sunit}{$E_{\rm h}^{-2}$\xspace}

\newcommand*{\cm}{cm$^{-1}$\xspace}
\newcommand*{\kcal}{kcal mol$^{-1}$\xspace}

\newcommand*{\PSI}{{\scshape Psi4}\xspace}
\newcommand{\mref}[0]{\Psi_0}
\newcommand{\tens}[3]{{#1}_{#2}^{#3}}

\newcommand{\cop}[1]{\hat{a}^{#1}}
\newcommand{\aop}[1]{\hat{a}_{#1}}
\newcommand{\sqop}[2]{\hat{a}_{#2}^{#1}}

\newcommand{\aphystei}[2]{\bra{#1}\!\!\ket{#2}}

\newcommand{\density}[2]{\gamma_{#2}^{#1}}

\newcommand{\no}[1]{ \{ {#1} \}}

\definecolor{goodorange}{RGB}{225,125,0}
\definecolor{goodgreen}{RGB}{5,130,5}
\definecolor{goodred}{RGB}{220,50,25}

\newcommand{\note}[2]{
\ifthenelse{\equal{#1}{F}}{
\colorbox{goodorange}{\textcolor{white}{\footnotesize \fontfamily{phv}\selectfont #1}}
    \textcolor{goodorange}{{\footnotesize \fontfamily{phv}\selectfont #2}}\xspace
}{}
\ifthenelse{\equal{#1}{Y}}{
\colorbox{goodred}{\textcolor{white}{\footnotesize \fontfamily{phv}\selectfont #1}}
    \textcolor{goodred}{{\footnotesize \fontfamily{phv}\selectfont #2}}\xspace
}{}
\ifthenelse{\equal{#1}{S}}{
\colorbox{goodgreen}{\textcolor{white}{\footnotesize \fontfamily{phv}\selectfont #1}}
    \textcolor{goodgreen}{{\footnotesize \fontfamily{phv}\selectfont #2}}\xspace
}{}
}

\emergencystretch=\maxdimen

\let\oldmaketitle\maketitle
\let\maketitle\relax

\title{Improving the efficiency of the multireference driven similarity renormalization group via sequential transformation, density fitting, and the non-interacting virtual orbital approximation}

\author{Tianyuan Zhang}
\email{tianyuan.zhang@emory.edu}
\affiliation{Department of Chemistry and Cherry L. Emerson Center for Scientific Computation, Emory University, Atlanta, Georgia 30322, USA}
\author{Chenyang Li}
\email{cli62@emory.edu}
\affiliation{Department of Chemistry and Cherry L. Emerson Center for Scientific Computation, Emory University, Atlanta, Georgia 30322, USA}
\author{Francesco A. Evangelista}
\email{francesco.evangelista@emory.edu}
\affiliation{Department of Chemistry and Cherry L. Emerson Center for Scientific Computation, Emory University, Atlanta, Georgia 30322, USA}
\begin{document}

\ifpreprint
\else
\twocolumn[
\begin{@twocolumnfalse}
\fi
\oldmaketitle

%
%
\begin{abstract}
This study examines several techniques to improve the efficiency of the linearized multireference driven similarity renormalization group truncated to one- and two-body operators \lbrack MR-LDSRG(2)\rbrack.
We propose a sequential MR-LDSRG(2) \lbrack sq-MR-LDSRG(2)\rbrack ~approach, in which one-body rotations are folded exactly into the Hamiltonian.
This new approach is combined with density fitting (DF) to reduce the storage cost of two-electron integrals. 
To further avoid the storage of large four-index intermediates, we propose a non-interacting virtual orbit (NIVO) approximation in which tensor blocks labeled by three and four virtual indices are neglected.
The NIVO approximation reduces the computational cost prefactor of the MR-LDSRG(2) bringing it closer to that of coupled cluster with singles and doubles (CCSD).
We test the effect of the DF and NIVO approximations on the MR-LDSRG(2) and sq-MR-LDSRG(2) methods by computing properties of eight diatomic molecules. The diatomic constants obtained by DF-sq-MR-LDSRG(2)+NIVO are found to be as accurate as those from the original MR-LDSRG(2) and coupled cluster theory with singles, doubles, and perturbative triples.
Finally, we demonstrate that the DF-sq-MR-LDSRG(2)+NIVO scheme can be applied to chemical systems with more than 550 basis functions by computing the automerization energy of cyclobutadiene with a quintuple-$\zeta$ basis set.
The predicted automerization energy is found similar to the values computed with Mukherjee's state-specific multireference coupled cluster theory with singles and doubles.

\end{abstract}

\ifpreprint
\else
\end{@twocolumnfalse}
]
\fi

\ifpreprint
\else
\small
\fi

\noindent





\section{Introduction}

The failure of conventional many-body methods to describe near-degenerate electronic states has motivated the development of many efficient and practical multireference approaches, including perturbation theories (MRPTs)\cite{Andersson:1990jz,*Andersson:1992cq,Hirao:1992bq,Nakano:1993hv,Werner:1996in,SinhaMahapatra:1999bp,Angeli:2001bg,*Angeli:2007by} and multireference configuration interaction (MRCI) schemes.\cite{Liu:1973fn,Werner:1988ku,*Knowles:1988hv,Hanrath:1997ch,Szalay:2012df,Sivalingam:2016hr}
Considerable efforts have been dedicate to the development of multireference coupled cluster (MRCC) methods,\cite{Cizek:1969hv,Jeziorski:1981gz,Haque:1984dk,Stolarczyk:1985fk,Malrieu:1985ce,Lindgren:1987in,Piecuch:1992gv,*Paldus:1993dx,Masik:1998gk,Mahapatra:1998cp,*Mahapatra:1998kj,*Mahapatra:1999ev,Li:2004ko,Hanrath:2005kj,Evangelista:2011eh,Hanauer:2011ey,Datta:2011ca,Chen:2012bm}
with the goal of creating nonperturbative theories that are both size extensive and systematically improvable.
Analogous many-body methods based on unitary transformations have received considerably less attention.\cite{Hoffmann:1988fs,Bartlett:1989iz,Watts:1989ht, Kutzelnigg:1991iw,Mertins:1996eu,Taube:2006bi,Chen:2012bm,Harsha:2018dv}
Unitary theories have, in principle, two advantages over traditional coupled cluster approaches: 1) the energy satisfies the variational principle, and 2) the transformed Hamiltonian is Hermitian.
The latter property is an important advantage in new applications of unitary methods to quantum computing, both in quantum algorithms\cite{Yung:2014iv,Peruzzo:2014kc,OMalley:2016dc,Shen:2017cc, Dumitrescu:2018fu, Ryabinkin:2018jw,Hempel:2018ip,Romero:2019hk, Lee:2018cy, Li:2019id, Kuhn:2018Accuracy, Nam:2019Ground} and downfolding approaches aimed at reducing the number of orbitals in quantum computations.\cite{Bauman:2019wj}

One of the main obstacles in the formulation of both single- and multi-reference unitary coupled cluster theories is that they lead to nonterminating equations.
The central quantity evaluated in these approaches is the similarity transformed Hamiltonian  ($\bar{H}$) defined as
\begin{equation}
\label{eq:unitarySTH}
\hat{H} \rightarrow \bar{H} = \hat{U}^\dagger \hat{H} \hat{U}= e^{-\hat{A}} \hat{H} e^{\hat{A}}
\end{equation}
where ($\hat{H}$) is the bare Hamiltonian and $\hat{U}$ is a unitary operator. In writing this transformation, we have expressed $\hat{U}$ as the exponential of the anti-Hermitian operator $\hat{A}$ ($\hat{A}^\dagger = - \hat{A}$), which is commonly written in terms of the coupled cluster excitation operator $\hat{T}$ as $\hat{A} = \hat{T} - \hat{T}^\dagger$.
Using the Baker--Campbell--Hausdorff (BCH) identity,\cite{Bartlett:1989iz,Kutzelnigg:1991iw,Taube:2006bi} the transformed Hamiltonian may be computed as the following commutator series
\begin{equation}
\label{eq:unitarycc}
\bar{H} = \hat{H} + [ \hat{H}, \hat{A}] + \frac{1}{2!}[ [ \hat{H}, \hat{A}], \hat{A}] + \frac{1}{3!}[ [ [ \hat{H}, \hat{A}], \hat{A}], \hat{A}]\ldots.
\end{equation}
Since the operator $\hat{A}$ contains both excitations and de-excitations, contractions are possible among components of $\hat{A}$, and as a consequence, the BCH series given in Eq.~\eqref{eq:unitarycc} is nonterminating.

Various approximations have been proposed to evaluate the unitarily transformed Hamiltonian.
Perhaps the simplest way to approximate the nonterminating unitary series is to truncate the BCH expansion after a certain number of commutators.\cite{Evangelista:2011jp,Hanauer:2012gf}
Proof-of-principle studies on unitary coupled cluster (CC) theory\cite{Evangelista:2011jp} suggest that for a a series containing up to $n$-nested commutators, the error decays as $10^{-n}$, and about four commutators are necessary to achieve sub-milliHartree accuracy.
Taube and Bartlett\cite{Taube:2006bi} have suggested tractable approximations to unitary CC theory based on the Zassenhaus expansion that are exact for a given number of electrons.
A common way to truncate the unitary BCH series is to use a recursive approximation of the commutator $[\,\cdot\,, \hat{A}]$, as suggested by Yanai and Chan.\cite{Yanai:2006gi,*Yanai:2007ix}
In their linear truncation scheme, these authors proposed to approximate each single commutator $[\,\cdot\,, \hat{A}]$ with its scalar and one- and two-body components, which we indicate as $[\,\cdot\,, \hat{A}]_{0,1,2}$.
Since in this truncation scheme the commutator $[\,\cdot\,, \hat{A}]_{0,1,2}$ preserves the many-body rank (number of creation and annihilation operators) of the Hamiltonian, the full BCH series can then be evaluated via a recursive relation.
An advantage of this approach is that closed-form expressions for terms like $[\hat{O}, \hat{A}]_{0,1,2}$, where $\hat{O}$ is an operator containing up to two-body terms  can be easily derived.
This truncation scheme has been employed in canonical transformation (CT) theory\cite{Yanai:2006gi,*Yanai:2007ix} and has been used to truncate normal-ordered  equations in the flow renormalization group of Wegner.\cite{Wegner:1994kh,Kehrein:2006vz}

We have recently developed a multireference driven similarity renormalization group (MR-DSRG)\cite{Evangelista:2014kt,Li:2015iz,Li:2016hb,*Li:2018dy,Li:2018kl,Li:2019xxx} approach that avoids the multiple-parentage problem\cite{Meller:1996ey,Mahapatra:1998cp,Hanrath:2005kj,Koehn:2013cp,Lyakh:2012cn,Evangelista:2018bt} and numerical instabilities\cite{Schucan:1972bs,*Schucan:1973dy,Salomonson:1980ko,Evangelisti:1987fw,Zarrabian:1990ig,Kowalski:2000cj,*Kowalski:2000cv,Lyakh:2012cn,Evangelista:2018bt} encountered in other nonperturbative multireference methods.
In the MR-DSRG, we perform a unitary transformation of the Hamiltonian controlled by a flow parameter, which determines to which extend the resulting effective Hamiltonian is  band diagonal.\cite{Li:2015iz,Li:2019xxx}
The simplest nonperturbative approximation, the linearized MR-DSRG truncated to two-body operators [MR-LDSRG(2)],\cite{Li:2016hb} assumes that $\hat{A}$ contains up to two-body operators (singles and doubles) and employs the linear commutator approximation of Yanai and Chan.
Preliminary benchmarks indicate that the MR-LDSRG(2) method is more accurate than CCSD around equilibrium geometries, and that this accuracy is preserved along  potential energy curves, especially for single-bond breaking processes.\cite{Li:2017bx,*Li:2018fn}
The cost to evaluate a single commutator in the MR-LDSRG(2) scales as ${\cal O} (N_{\bf C}^2  N_{\bf V}^2 N^2) = {\cal O} (N_{\bf C}^2 N_{\bf V}^4 + N_{\bf C}^3 N_{\bf V}^3 + \ldots)$ where $N_{\bf C}$, $N_{\bf V}$, and $N$ are the numbers of core, virtual, and total orbitals, respectively.
This scaling is identical to that of CC with singles and doubles (CCSD) [${\cal O} (N_{\bf C}^2 N_{\bf V}^4)$]. However, while in CCSD the most expensive term is evaluated only once, in the MR-LDSRG(2) the same term must be evaluated for each nested commutator.
Consequently, if the BCH series is truncated after $n + 1$ terms, the computational cost of the MR-LDSRG(2) is roughly $n$ times that of CCSD, where approximately ten or more terms are usually required to convergence the energy error in the BCH series to $10^{-12}$ \Eh.\cite{Li:2016hb}
A second reason is that computing the BCH series requires storing large intermediate tensors with memory costs that scale as $\mathcal{O}(N^4)$.
When these intermediates are stored in memory, practical computations are limited to 200--300 basis functions on a single computer node.

In this work, we combine a series of improvements and approximations to reduce the computational and memory requirements of the MR-LDSRG(2) down to a small multiple of the cost of CCSD.
To begin, we consider an alternative ansatz for the MR-DSRG based on a sequential similarity transformation.\cite{Evangelista:2012fo}
This sequential ansatz reduces the complexity of the MR-DSRG equations and, when combined with integral factorization techniques, reduces significantly the cost to evaluate singles contributions.
Second, we apply density fitting (DF)\cite{Whitten:1973ju,Dunlap:1979gh,Kendall:1997kh} to reduce the memory requirements and the I/O cost by avoiding the storage of of two-electron repulsion integrals.
Together with Cholesky decomposition (CD)\cite{Beebe:1977dp,Roeggen:1986fp,Koch:2003go,Boman:2008fw,Aquilante:2009eu} and other tensor decomposition schemes,\cite{Parrish:2014ig} these techniques have been crucial in enabling computations with 1000 or more basis functions and found application in numerous electronic structure methods,\cite{Vahtras:1993db,Aquilante:2007es,Feyereisen:1993ja,Weigend:2002ic,Werner:2003gq,Aquilante:2008gk,Bostrom:2010is,Gyorffy:2013kf,Hannon:2016bh,Freitag:2017ir,Hattig:2000cr,Pedersen:2004kn}
including coupled cluster methods.\cite{Rendell:1994kl,Bostrom:2012gq,DePrince:2013ki,Epifanovsky:2013gd,Qiu:2017in}
Third, we reduce the cost of MR-LDSRG(2) computations by neglecting operators that involve three or more virtual electrons.
We term this truncation scheme the non-interacting virtual orbital (NIVO) approximation.
A perturbative analysis of the NIVO approximation shows that the errors introduced appear at third order.
To the best of our knowledge, such approximation has not been introduced in multireference theories, but it is analogous to other truncation schemes used in CCSD in which certain diagrams are have modified coefficients\cite{Huntington:2010ef,Evangelista:2012hz,Kats:2013jy,*Kats:2014bg,Rishi:2017ky} or are completely removed.\cite{Koch:1981dx,Bartlett:2006fh}

This paper is organized in the following way.
In Sec.~\ref{sec:theory} we present an overview of the MR-DSRG theory, discuss the sequential MR-DSRG, and introduced the NIVO approximation.
Details of the implementation together with a discussion of timings are given in Sec.~\ref{sec:impl}.
In Sec.~\ref{sec:result} we assess the accuracy of several MR-LDSRG(2) schemes on a benchmark set of several diatomic molecules and determine the automerization barrier of cyclobutadiene.
Finally, in Sec.~\ref{sec:conclusion} we conclude this work with a discussion of the main results.

\section{Theory}
\label{sec:theory}

We first define the orbital labeling convention employed in this work.
The set of molecular spin orbitals ${\bf G} \equiv \{\phi^p, p = 1, 2, \dots, N\}$ is partitioned into core (\textbf{C}), active (\textbf{A}), and virtual (\textbf{V}) components of sizes $N_{\bf C}$, $N_{\bf A}$, and $N_{\bf V}$, respectively.
We use indices $m,n$ to label core orbitals, $u,v,x,y$ to label active orbitals, and $e,f,g,h$ to label virtual orbitals.
For convenience, we also define the set of hole ($\bf H = C \cup A$) and particle ($\bf P = A \cup V$) orbitals with dimensions $N_{\bf H} = N_{\bf C} + N_{\bf A}$ and $N_{\bf P} = N_{\bf A} + N_{\bf V}$, respectively.
Hole orbitals are denoted by indices $i,j,k,l$ and those of particle by indices $a,b,c,d$.
General orbitals are labeled by indices $p,q,r,s$.

The MR-DSRG assumes a multideterminantal reference wave function ($\mref$):
\begin{equation}
\label{eq:ref}
\ket{\mref} = \sum_{\mu = 1}^{d} c_{\mu} \ket{\Phi^{\mu}}.
\end{equation}
In this work, the set of determinants $\{\Phi^{\mu}; \mu = 1, 2, \dots, d\}$ in Eq.~\eqref{eq:ref} is assumed to form a complete active space (CAS) and the corresponding coefficients $\{c_{\mu}; \mu = 1, 2, \dots, d\}$ are obtained from a CAS self-consistent field (CASSCF) computation.\cite{Roos:1980fd} 
All operators are then normal ordered with respect to $\mref$ according to the scheme of Mukherjee and Kutzelnigg.\cite{Mukherjee:1997kf,*Kutzelnigg:1997dp,*Kutzelnigg:2010iu,*Kong:2010kg,*Sinha:2013dx}
For example, the bare Hamiltonian is expressed as
\begin{equation}
\label{eq:bareH}
\hat{H} = E_0 + \sum_{pq} f_p^q \no{\sqop{p}{q}} + \frac{1}{4} \sum_{pqrs} v_{pq}^{rs} \no{\sqop{pq}{rs}} ,
\end{equation}
where $E_0=\braket{\mref|\hat{H}|\mref}$ is the reference energy and $\no{\sqop{pq\dots}{rs\dots}} = \no{ \cop{p} \cop{q} \dots \aop{s} \aop{r} }$ is a product of creation ($\cop{p} \equiv \aop{p}^\dag$) and annihilation ($\aop{p}$) operators in its normal-ordered form, as indicated by the curly braces.
The generalized Fock matrix ($f_p^q$) introduced in Eq.~\eqref{eq:bareH} is defined as
\begin{equation}
\label{eq:fock}
f_p^q = h_p^q + \sum_{rs} \tens{v}{pr}{qs} \density{r}{s},
\end{equation}
where $\tens{h}{p}{q} = \braket{\phi_p | \hat{h} | \phi^q}$ and $\tens{v}{pq}{rs} = \aphystei{\phi_p \phi_q}{\phi^r \phi^s}$ are the one-electron and anti-symmetrized two-electron integrals, respectively.
Here, we have also used the one-particle reduced density matrix (1-RDM) defined as $\density{p}{q} = \braket{\mref| \sqop{p}{q} |\mref}$.

\subsection{Review of the MR-DSRG method}
\label{subsec:mrdsrg}
  
The MR-DSRG performs a parametric unitary transformation of the bare Hamiltonian analogous to Eq.~\eqref{eq:unitarySTH}, whereby the anti-Hermitian operator $\hat{A} (s)$ depends on the so-called flow parameter, a real number $s$ defined in the range of $[0,\infty)$. The resulting transformed Hamiltonian [$\bar{H}(s)$] is a function of $s$ defined as
\begin{align}
\label{eq:dsrg}
\bar{H} (s) = e^{-\hat{A} (s)} \hat{H} e^{\hat{A} (s)}.
\end{align}
The operator $\hat{A}(s)$ is a sum of many-body operators with rank ranging from one up to the total number of electrons ($n$),
\begin{align}
\label{eq:A_mb}
\hat{A}(s) = \sum_{k=1}^{n} \hat{A}_k(s),
\end{align}
where $\hat{A}_k(s)$ is the $k$-body component of $\hat{A}(s)$.
In the DSRG the operators $\hat{A}_k (s)$ are written as $\hat{A}_k(s) = \hat{T}_k (s) - \hat{T}^\dag_k (s)$, where $\hat{T}_k (s)$ is an $s$-dependent cluster operator defined as
\begin{equation}
\label{eq:T_k}
\hat{T}_k(s) = \frac{1}{(k!)^2} \sum_{ij\cdots}^{\mathbf{H}} \sum_{ab\cdots}^{\mathbf{P}} \tens{t}{ab\cdots}{ij\cdots} (s) \no{\sqop{ab\cdots}{ij\cdots}}.
\end{equation}
Note that the cluster amplitudes $\tens{t}{ab\cdots}{ij\cdots} (s)$ exclude internal excitations, which are labeled only with active orbital indices.
The DSRG transformed Hamiltonian has a many-body expansion similar to Eq.~\eqref{eq:bareH},
\begin{equation}
\bar{H}(s)=\bar{E}_0(s)+\sum_{pq} \bar{H}_q^p(s)\{\hat{a}_p^q\}+\frac{1}{4}\sum_{pqrs} \bar{H}_{rs}^{pq}(s) \{\hat{a}_{pq}^{rs}\}+\cdots,
\end{equation}
where,
\begin{equation}
\label{eq:Eu}
\bar{E}_0(s) = \braket{\mref| \bar{H}(s) |\mref},
\end{equation}
is the DSRG energy and the tensors $\bar{H}_{rs\dots}^{pq\dots}(s)$ are analogous to one- and two-electron integrals but dressed with dynamical correlation effects.

The goal of the DSRG transformation is to decouple the interactions between the reference wave function ($\mref$) and its excited configurations.
Such interactions are the couplings between hole and particle orbitals represented by generalized excitation [$\tens{\bar{H}}{ab\dots}{ij\dots} (s) \no{\sqop{ab\dots}{ij\dots}}$] and de-excitation [$\tens{\bar{H}}{ij\dots}{ab\dots} (s) \no{\sqop{ij\dots}{ab\dots}}$] operators, where $ij\cdots \in \mathbf{H}$ and $ab\cdots \in \mathbf{P}$, excluding cases where all the indices are active orbitals.
These terms of $\bar{H}(s)$ that the DSRG transformation aims to suppress are called the off-diagonal components and will be denoted as $\bar{H}_{\rm od}(s)$.
Instead of achieving a full decoupling of the off-diagonal components [i.e., $\bar{H}_{\rm od}(s) = 0$], we demand that the DSRG transformation achieves a partial decoupling, avoiding the components of $\bar{H}_{\rm od}(s)$ with small or vanishing M{\o}ller--Plesset energy denominators. 
This condition is imposed via the DSRG flow equation, a nonlinear implicit equation of the form
\begin{equation}
\label{eq:dsrg_flow}
\bar{H}_{\rm od}(s) = \hat{R} (s),
\end{equation}
where the \emph{source operator} $\hat{R}(s)$ is Hermitian and continuous in $s$.
The role of the source operator in the DSRG flow equation is to drive the off-diagonal elements of $\bar{H}(s)$ to zero, a goal achieved by an appropriate parameterization of $\hat{R}(s)$.\cite{Evangelista:2014kt}

In the linearized MR-DSRG approximation [MR-LDSRG(2)], the BCH series [Eq.~\eqref{eq:unitarycc}] is evaluated by keeping up to two-body normal-ordered operators of each commutator. The transformed Hamiltonian can then be evaluated by the following recursive equations
\begin{equation}
\label{eq:recursive}
\left\{
\begin{split}
\hat{C}^{(k+1)}(s) =&\, \frac{1}{k + 1}[\hat{C}^{(k)}_{1,2} (s), \hat{A}(s)]_{0,1,2},\\
\bar{H}^{(k+1)}(s) =&\, \bar{H}^{(k)}(s) +  \hat{C}^{(k+1)} (s),
\end{split}\right.
\end{equation}
starting from $\hat{C}^{(0)}(s) = \bar{H}^{(0)} = \hat{H}$ and iterating until the norm of $\hat{C}^{(k+1)}_{1,2} (s)$ is less than a given convergence threshold.

The solution of Eq.~\eqref{eq:dsrg_flow} yields a set of amplitudes $\tens{t}{ab\dots}{ij\dots} (s)$ that define the operator $\hat{A}(s)$ and the DSRG transformed Hamiltonian $\bar{H}(s)$.
From this latter quantity, the MR-DSRG electronic energy is computed as the expectation value with respect to the reference
\begin{equation}
\label{eq:Eu}
\bar{E}_0(s) = \braket{\mref| \bar{H}(s) |\mref}.
\end{equation}
We refer the energy computed using Eq.~\eqref{eq:Eu} as the \textit{unrelaxed} energy since the reference coefficients are not optimized.
To include reference relaxation effects, we require that $\mref$ is an eigenstate of $\bar{H}(s)$ within the space of reference determinants, a condition that is equivalent to solving the eigenvalue problem
\begin{equation}
\label{eq:eigen}
    \sum_{\mu=1}^{d} \braket{\Phi_{\nu} | \bar{H}(s) | \Phi^{\mu}} c'_{\mu} = E(s) c'_{\nu}.
\end{equation}
Equation \eqref{eq:eigen} defines a new reference $\mref'$ with expansion coefficients $c'_{\mu}$.
This new reference may be used as a starting point for a subsequent MR-DSRG transformation and this procedure can be repeated until convergence and such converged energy is referred as \textit{fully relaxed} energy.
For the nonperturbative MR-DSRG schemes discussed in this work, we use the fully relaxed energy by default, unless otherwise noted.

\subsection{Simplifying the MR-DSRG equations: Sequential transformation}
\label{subsec:seq}



Our first modification to the MR-DSRG approach is an alternative way to transform the bare Hamiltonian via a sequence of unitary operators with increasing particle rank
\begin{equation}
\label{eq:dsrg_sq}
\bar{H}(s) = e^{-\hat{A}_n(s)} \cdots e^{-\hat{A}_2(s)} e^{-\hat{A}_1(s)} \hat{H} e^{\hat{A}_1(s)} e^{\hat{A}_2(s)} \cdots e^{\hat{A}_n(s)}.
\end{equation}
We term the MR-DSRG approach based on Eq.~\eqref{eq:dsrg_sq} the \textit{sequential} MR-DSRG (sq-MR-DSRG), while we refer to the original formalism based on Eq.~\eqref{eq:dsrg} as the \textit{traditional} MR-DSRG.
Note that in the limit of $s \rightarrow \infty$ and no truncation of $\hat{A}(s)$, both the traditional and sequential MR-DSRG can approach the full configuration interaction limit.\cite{Evangelista:2012fo}
However, these schemes are not equivalent for truncated $\hat{A} (s)$ [for example, $n = 2$ in Eqs.~\eqref{eq:A_mb} and \eqref{eq:dsrg_sq}] due to the fact that operators of different particle rank do not commute, that is, $[\hat{A}_i (s), \hat{A}_j (s)] \neq 0$ for $i \neq j$.

An advantage of the sq-MR-DSRG approach is that $\hat{A}_1 (s)$ can be exactly folded into the Hamiltonian via a unitary transformation.
The resulting $\hat{A}_1 (s)$-dressed Hamiltonian $[\tilde{H}(s)]$,
\begin{equation}
\label{eq:dressedH}
\tilde{H}(s) = e^{-\hat{A}_1(s)} \hat{H} e^{\hat{A}_1(s)},
\end{equation}
preserves the particle rank of the bare Hamiltonian [Eq.~\eqref{eq:bareH}]. The corresponding scalar and tensor components of $\tilde{H}(s)$ can be obtained by a simple unitary transformation of the one- and two-electron integrals ($f_p^q$ and $v_{pq}^{rs}$) and update of the scalar energy.
As will be discussed in Sec.~\ref{sec:impl_sq}, the $\hat{A}_1 (s)$-dressed Hamiltonian can be computed very efficiently when the two-electron integrals are approximated with DF or CD.

The transformed Hamiltonian for the sq-MR-DSRG truncated to one- and two-body operators is given by
\begin{equation}
\label{eq:sqHbar}
\bar{H}(s)  = e^{-\hat{A}_2(s)} \tilde{H}(s) e^{\hat{A}_2(s)}.
\end{equation}
In the linear approximation, the evaluation of Eq.~\eqref{eq:sqHbar} is simpler than in the traditional MR-LDSRG(2) since the total number of tensor contractions is reduced from 39 to 30.\cite{Li:2016hb}
Another advantage of the sequential approach is that $\hat{A}_1 (s)$ is treated exactly, while in the traditional scheme some contractions involving singles are neglected.
To appreciate this point, consider all the contributions to the double-commutator term in the MR-LDSRG(2) that depend on $\hat{A}_1(s)$
\begin{equation}
\label{eq:traditional_two_comm}
\begin{split}
[[\hat{H}, \hat{A}(s)]_{1,2}, \hat{A}(s)]_{0,1,2} \leftarrow ~~&  [[\hat{H}, \hat{A}_1(s)]_{1,2}, \hat{A}_1(s)]_{0,1,2}\\
+& [[\hat{H}, \hat{A}_1(s)]_{1,2}, \hat{A}_2(s)]_{0,1,2}\\
+& [[\hat{H}, \hat{A}_2(s)]_{1,2}, \hat{A}_1(s)]_{0,1,2}.
\end{split}
\end{equation}
The first term on the r.h.s.\ of Eq.~\eqref{eq:traditional_two_comm} is treated exactly in the MR-LDSRG(2).
However, since contractions involving $\hat{A}_2(s)$ generate three-body terms  (truncated in the linearized approximation), the contribution of $\hat{A}_1(s)$ in the second and third terms are not included exactly in the MR-LDSRG(2) transformed Hamiltonian.
In the sequential approach, all contributions from $\hat{A}_1(s)$ are treated by forming the operator $\tilde{H}(s)$, and since the BCH expansion for such transformation does not generate intermediates with rank greater than two, all terms involving $\hat{A}_1(s)$ are treated exactly in the linearized approximation.

\subsection{Alleviating the memory bottleneck: The non-interacting virtual orbital (NIVO) approximation}
\label{subsec:NIVO}

In both the traditional and sequential MR-DSRG approaches, the DF approximation reduces the cost to store both the bare and $\hat{A}_1 (s)$-dressed Hamiltonian from ${\cal O}(N^4)$ to ${\cal O}(N^2M)$.
However, in the evaluation of the recursive commutator approximation of $\bar{H}(s)$, two-body operators are generated during the evaluation of each intermediate commutator [$\hat{C}^{(k)} (s)$] and $\bar{H}(s)$.
These quantities have ${\cal O}(N^4)$ storage cost and, thus, reintroduce the bottleneck avoided with DF.

In order to reduce the cost to store $\bar{H}_{2}(s)$ and $\hat{C}^{(k)}_2 (s)$, we shall neglect certain tensor blocks of these operators.
By partitioning of orbitals into core ($\mathbf{C}$), active ($\mathbf{A}$) and virtual ($\mathbf{V}$) spaces, each general 4-index tensor may be subdivided into 81 blocks according to the combination of orbital indices, for example $\mathbf{CCCC}$, $\mathbf{AAVV}$, $\mathbf{CAVA}$, etc.
We propose a non-interacting virtual orbital (NIVO) approximation, which neglects the operator components of $\hat{C}^{(k+1)}_{2} (s) = [\hat{C}^{(k)}_{1,2} (s),\hat{A}(s)]_2$, $k \geq 0$, with three or more virtual orbital indices ($\mathbf{VVVV}$, $\mathbf{VCVV}$, $\mathbf{VVVA}$, etc.) in the recursive definition of the linearized BCH series.
Neglecting these blocks, the number of elements in each NIVO-approximated tensor is reduced from ${\cal O}(N^4)$ to ${\cal O}(N^2N_\mathbf{H}^2)$, a size comparable to that of the $\hat{A}_2(s)$ tensor.
For instance, in the cyclobutadiene computation using a quintuple-$\zeta$ basis set reported in Sec.~\ref{subsec:cyclobutadiene}, the memory requirements of $\bar{H}_{2}(s)$ or $\hat{C}^{(k)}_2 (s)$ are reduced from 2.7 TB to 6.8 GB by the NIVO approximation.

\begin{figure}[!h]
\ifpreprint
    \includegraphics[width=0.80\columnwidth]{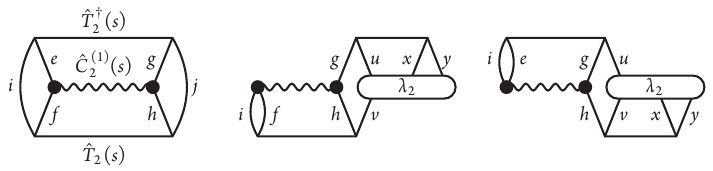}
\else
    \includegraphics[width=\columnwidth]{ignored_diagrams.pdf}
\fi
\caption{Diagrams that are neglected by the NIVO approximation in the evaluation of the term $\hat{C}_{0}^{(3)} (s)$ as in Eq.~\eqref{eq:nivo_error}. The wiggly and horizontal solid lines indicate the effective interaction of $\hat{C}_2^{(1)} (s)$ and $\hat{A}_2(s)$, respectively. The two-body density cumulant is labeled by $\lambda_2$.}
\label{fig:diagram}
\end{figure}

To justify the NIVO approximation we analyze its effect on the energy.
The first term in the BCH series that is approximated in the sq-MR-LDSRG(2)+NIVO scheme is the commutator $\hat{C}_{2}^{(1)}(s) = [\tilde{H} (s), \hat{A}_2 (s)]_{2}$.
Indicating the terms neglected in NIVO as $\delta \hat{C}_{2}^{(1)}(s)$, we see that the first energy contribution affected by the NIVO approximation comes from the expectation value of the triple commutator term [$\delta \hat{C}_{0}^{(3)} (s)$]
\begin{equation}
\label{eq:nivo_error}
\delta \hat{C}_{0}^{(3)} (s) = \frac{1}{6}[[\delta \hat{C}_{2}^{(1)}(s),  \hat{A}_2(s)]_{1,2}, \hat{A}_2(s)]_{0},
\end{equation}
whose contributions are shown as diagrams in Fig.~\ref{fig:diagram}.
From a perturbation theory perspective, these diagrams are of order four or higher [assuming $\hat{A}_2(s)$ to be of order one] and, therefore, are negligible compared to the leading error (third order) of the linearized commutator approximation.

Hereafter, we shall append ``+NIVO'' at the end of the method name to indicate the use of NIVO approximation.
For example, the density fitted MR-LDSRG(2) method in the sequential transformation ansatz with NIVO approximation is termed ``DF-sq-MR-LDSRG(2)+NIVO''.

\section{Implementation}
\label{sec:impl}

The sq-MR-LDSRG(2) method combined with DF and the NIVO approximation was implemented in \textsc{Forte},\cite{FORTE2019} an open-source suite of multireference theories for molecular computations.
This implementation reuses several components of our previous MR-LDSRG(2) code based on conventional four-index two-electron integrals.\cite{Li:2016hb}
The DSRG equations were implemented as tensor contractions using the \textsc{Ambit} tensor library,\cite{AMBIT2018} while integrals were generated using the \textsc{Psi4} package.\cite{Parrish:2017hg,*Smith:2018ip}
In the following, we provide the details of our implementation of the sequential ansatz in combination with DF.\cite{Hohenstein:2012hk}

\subsection{Sequential transformation}
\label{sec:impl_sq}

The $\hat{A}_1 (s)$-dressed Hamiltonian [$\tilde{H}(s)$, Eq.~\eqref{eq:dressedH}] can be obtained by a unitary transformation of $\hat{H}$ via the operator $\hat{U} (s) = \exp[\hat{A}_1(s)]$.
For clarity, we shall drop the label ``(s)'' for all $s$-dependent quantities [$\tilde{H} (s), \hat{A}_1(s)$, and $\hat{U} (s)$] in this section.
The one- and two-body components of $\tilde{H}$ ($\tilde{h}^{q'}_{p'}$ and $\tilde{v}_{p'q'}^{r's'}$) are given by
\begin{align}
\label{eq:h_trans} 
\tilde{h}^{q'}_{p'} &= \sum_{pq} U_{q'}^q\, h^q_p\,U_p^{p'},\\ 
\label{eq:v_trans}
\tilde{v}_{p'q'}^{r's'} &= \sum_{pqrs} U_{r}^{r'}\,U_{s}^{s'}\,v_{pq}^{rs} \,U_{p'}^{p}\,U_{q'}^{q}.
\end{align}
Here, the unitary matrix $\tens{U}{p}{p'}=(\mathbb{U})_{p'p}$ and its inverse $\tens{U}{p'}{p} = (\mathbb{U})_{p'p}^\ast$ are given by $\mathbb{U} = e^{\mathbb{A}}$, where the matrix $\mathbb{A}$ is composed of elements of the $\hat{A}_1$ tensor, 
 $(\mathbb{A})_{ia}=\tens{t}{a}{i}$ and $(\mathbb{A})_{ai}=-\tens{t}{a}{i}$.
Note that we use primed indices only as a way to distinguish labels, yet these indices by no means imply a new set of orbitals.

The $\hat{A}_1$-dressed Hamiltonian written in normal ordered form with respect to $\mref$ is given by
\begin{equation}
\label{eq:dressedH_struct}
\tilde{H} = \tilde{E}_0 + \sum_{pq}\tilde{f}_p^q\{\hat{a}_q^p\} + \frac{1}{4} \sum_{pqrs} \tilde{v}_{pq}^{rs} \{\hat{a}_{rs}^{pq}\},
\end{equation}
where the transformed energy ($\tilde{E}_0$) is given by 
\begin{equation}
\label{eq:E0_sq_trans}
\tilde{E}_0 = \sum_{i'j'}^{\mathbf{H}} \tilde{h}_{i'}^{j'} \gamma_{j'}^{i'} + \frac{1}{4}\sum_{i'j'k'l'}^{\mathbf{H}}\tilde{v}^{k'l'}_{i'j'}\gamma^{i'j'}_{k'l'},
\end{equation}
and the Fock matrix elements ($\tilde{f}^{q'}_{p'}$) are defined as
\begin{equation}
\tilde{f}^{q'}_{p'} = \tilde{h}^{q'}_{p'} + \sum_{i'j'}^{\mathbf{H}}\tilde{v}^{q'j'}_{p'i'} \density{i'}{j'}.\label{eq:f_trans}
\end{equation}
Note that the quantities $\density{i'}{j'}$ and $\density{i'j'}{k'l'}$ in Eqs.~\eqref{eq:E0_sq_trans} and \eqref{eq:f_trans} are the \textit{untransformed} 1- and 2-RDMs of the reference $\ket{\mref}$ defined as $\density{p}{q} = \braket{\mref| \sqop{p}{q} |\mref}$ and $\density{pq}{rs} = \braket{\mref| \sqop{pq}{rs} |\mref}$, respectively.

The two-electron integral transformation [Eq.~\eqref{eq:v_trans}] has a noticeable cost [${\cal O}(N^5)$] and must be repeated each time the $\hat{A}_1$ operator is updated.
However, in the implementation based on DF integrals, this transformation may be performed in a significantly more efficient way.
In DF, the four-index electron repulsion integral tensor as a contraction of a three-index auxiliary tensor ($B_{pq}^{Q}$),
\begin{equation}
\label{eq:int_fact}
    \braket{pq | rs} \approx \sum_Q^M B^Q_{pr} B^{Q}_{qs},
\end{equation}
where $M$ is the dimension of the fitting basis in DF.
Using this decomposition, the unitary transformation may be directly applied to each auxiliary tensor,
\begin{equation}
\label{eq:B_trans}
\tilde{B}^{Q}_{p'q'} = \sum_{pq}  B^{Q}_{pq}\, U^{p}_{p'}\,U^{q}_{q'},
\end{equation}
reducing the cost to evaluate $\tilde{H}$ down to ${\cal O} (N^3 M)$.

Equations \eqref{eq:dressedH_struct}--\eqref{eq:E0_sq_trans} specify the procedures to obtain $\tilde{H}$ as a unitary transformation of $\hat{H}$.
Since $\tilde{H}$ retains the structure of $\hat{H}$, we can reuse most of our previous MR-LDSRG(2) code\cite{Li:2016hb} to implement sq-MR-LDSRG(2) by employing $\tilde{H}$ (instead of $\hat{H}$) and removing terms involving $\hat{A}_1$.

As described in Ref.~\citenum{Li:2016hb}, we evaluate the  commutator $\hat{C}^{(k+1)} = \frac{1}{k + 1}[\hat{C}^{(k)}_{1,2} , \hat{A}]_{0,1,2}$ in Eq.~\eqref{eq:recursive} using the following recursive system of equations

since $[\hat{C}^{(k)}_{1,2} , \hat{T}^\dagger] = - [\hat{C}^{(k)}_{1,2} , \hat{T}]^\dagger$, computing the
\begin{equation}
\label{eq:unitary_comm}
\left\{
\begin{split}
\hat{O}^{(k+1)} & = \frac{1}{k + 1}[\hat{C}^{(k)}_{1,2} , \hat{T}]_{0,1,2},\\
\hat{C}^{(k+1)} &= \hat{O}^{(k+1)} +  [\hat{O}^{(k+1)}]^\dagger,
\end{split}\right.
\end{equation}
where $\hat{O}^{(k+1)}$ is an intermediate containing up to two-body components.
The iteration starts from either $\hat{C}^{(0)} = \hat{H}$, in traditional MR-LDSRG(2), or $\hat{C}^{(0)} = \tilde{H}$ in the sequential version, optionally applying the NIVO approximation to the two-body intermediate tensors $\hat{O}^{(k)}_{2}$, $\hat{C}^{(k)}_{2}$ and $\bar{H}^{(k)}_{2}$ for $k \geq 1$.

\subsection{Batched tensor contraction for the DF algorithm}
\label{sec:batch}

Despite the storage cost reduction of the DF and NIVO approximations, another potential memory bottleneck is the size of the intermediate tensors generated during the evaluation of commutators.
For example, consider the following contraction, 
\begin{equation}
\label{eq:eri_contraction}
\tens{O}{rs}{ij} \leftarrow  \sum_{ab}^{\mathbf{P}} \aphystei{rs}{ab} \tens{t}{ab}{ij} \quad \forall i,j \in \mathbf{H}, \forall r,s \in \mathbf{G},
\end{equation}
which is also found in the CCSD equations.
In the DF case, Eq.~\eqref{eq:eri_contraction} is written as two contractions involving auxiliary tensors,
\begin{equation}
\label{eq:df_contraction}
\begin{split}
\tens{O}{rs}{ij} \leftarrow & \sum_{ab}^{\mathbf{P}} \sum_{Q}^{M} B^{Q}_{ar} B^{Q}_{bs} \tens{t}{ab}{ij}
  - \sum_{ab}^{\mathbf{P}} \sum_{Q}^{M} B^{Q}_{as} B^{Q}_{br} \tens{t}{ab}{ij}.
\end{split}
\end{equation}
The most efficient way to evaluate the first term of Eq.~\eqref{eq:df_contraction} is to introduce the intermediate tensor  $I_{arbs} = \sum_{Q}^{M} B^{Q}_{ar} B^{Q}_{bs}$ of size ${\cal O}(N^2 N_\mathbf{P}^2)$.
To avoid storage of these large intermediates, it is common to evaluate Eq.~\eqref{eq:df_contraction} using a batched algorithm, whereby a slice of the tensor $I_{arbs}$ is computed and contracted on the fly with the amplitudes $\tens{t}{ab}{ij}$.
To automate this optimization of the tensor contraction we have coded a generic batching algorithm in the tensor library \textsc{Ambit}.\cite{AMBIT2018}
Whereas the \textsc{Ambit} code for the first term in Eq.~\eqref{eq:df_contraction} is written as
\begin{Verbatim}[fontsize=\footnotesize]
O["ijrs"] += B["Qar"] * B["Qbs"] * t["ijab"];
\end{Verbatim}
our new implementation allows batching over the index $r$ by simply surrounding the contraction with the \verb|batched()| function decorator
\begin{Verbatim}[fontsize=\footnotesize]
O["ijrs"] += batched("r", B["Qar"] * B["Qbs"] * t["ijab"]);
\end{Verbatim}

\begin{figure}
\begin{algorithm}[H]
\caption{The batched algorithm to compute $\tens{C}{rs}{ij} \leftarrow \sum_{ab}^{\mathbf{P}} \sum_{Q}^{M} B^{Q}_{ar} B^{Q}_{bs} \tens{t}{ab}{ij}$.}
\label{alg:batch}
\begin{algorithmic}[1]
\State Permute memory layout of $\tens{C}{rs}{ij}$ and $B^{Q}_{ar}$ so that their $r$-subblocks $\tens{C}{[r]s}{ij}$ and $B^{Q}_{a[r]}$ are contiguous in memory.
\For{ each $r = 1, 2, \ldots, N_\mathbf{G}$}
\State $I_{abs} := \sum_{Q}^{M} B^{Q}_{a[r]} B^{Q}_{bs}$
\State $\tens{C}{[r]s}{ij} \leftarrow  \sum_{ab}^{\mathbf{P}} I_{abs} \tens{t}{ab}{ij} \quad i,j \in \mathbf{H}, r \in \mathbf{G}$
\EndFor
\State Permute $\tens{C}{rs}{ij}$ back to the original memory layout.
\end{algorithmic}
\end{algorithm}
\end{figure}

\subsection{Computational cost reduction}
\label{subsec:timing}

\begin{figure}[!htp]
\ifpreprint
    \includegraphics[width=0.5\columnwidth]{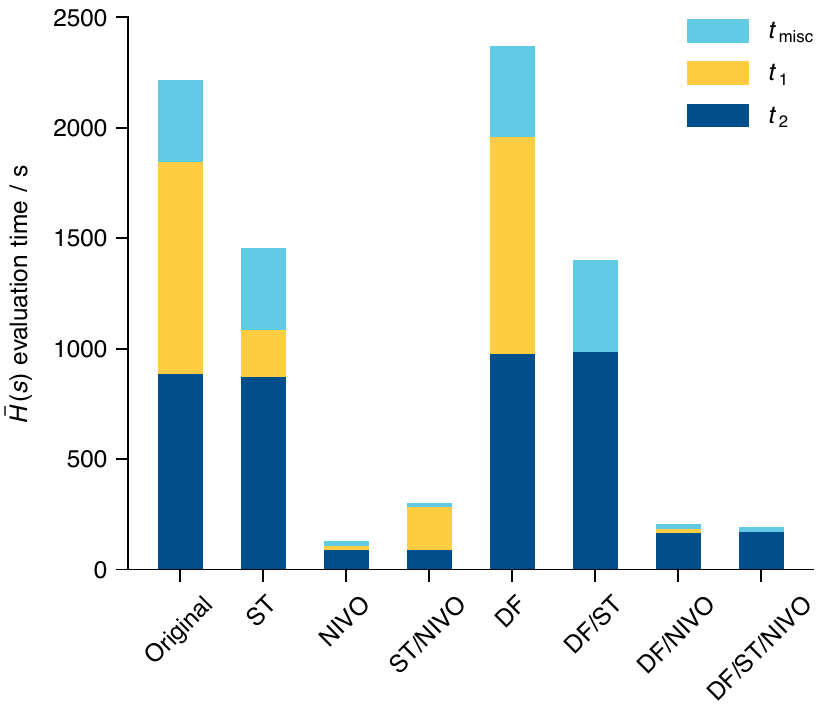}
\else
    \includegraphics[width=0.85\columnwidth]{C4H4_TZ_timing.pdf}
\fi
\caption{The time to evaluate the DSRG transformed Hamiltonian [$\bar{H}$] of the ground-state cyclobutadiene when different techniques are introduced to the MR-LDSRG(2) method. These techniques include: density fitting (DF), sequential transformation (ST), and the non-interacting virtual orbital (NIVO) approximation. The total time of computing $[\hat{C}^{(k)} , \hat{T}_1 ]$ in MR-LDSRG(2) or $\tilde{H} = e^{-\hat{A}_1} \hat{H} e^{\hat{A}_1}$ in sq-MR-LDSRG(2) is labeled as $t_1$ in this plot. All computations employed the cc-pVTZ basis set and they were carried out on an Intel Xeon E5-2650 v2 processor using 8 threads.}
\label{fig:timing}
\end{figure}

Here we discuss timings for all the MR-LDSRG(2) variants introduced in this work.
In MR-LDSRG(2) theory, the computational bottleneck is forming the DSRG transformed Hamiltonian $\bar{H}$.
Timings for computing $\bar{H} $ in the case of cyclobutadiene (see Sec.~\ref{subsec:cyclobutadiene} for details) are summarized in Fig.~\ref{fig:timing}.
Detailed timings for the evaluation of $\bar{H}$ using all the combinations of the approximations considered here are reported in the Supplementary Material.

The total timing ($t_\mathrm{tot}$) for computing the transformed Hamiltonian in $n$ iterations is partitioned according to 
\begin{equation}
t_\mathrm{tot} = t_1 + t_2 + t_\mathrm{misc},
\end{equation}
where $t_1$ and $t_2$ are the timings to evaluate the commutators involving $\hat{T}_1$ and $\hat{T}_2$, respectively. In the sequential transformation approach, $t_1$ is instead defined as the time for forming the $\hat{A}_1$-transformed Hamiltonian.
The term $t_\mathrm{misc}$ accounts for the cost to sort and accumulate the results of contractions with $\hat{T}_1$ and $\hat{T}_2$, as shown in the second line of Eq.~\eqref{eq:unitary_comm}.
Figure~\ref{fig:timing} shows that the timing for the conventional MR-LDSRG(2) is dominated by contractions involving $\hat{T}_1$ and $\hat{T}_2$.
The cost of the singles contractions can be reduced significantly (3--5 times) by employing the sequentially transformed approach, even though at each iteration of the sq-MR-LDSRG(2) equations it is necessary to build the operator $\tilde{H}$.

Applying the NIVO approximation to the original MR-LDSRG(2) leads to a drastic reduction of the total computational time ($\times$18 speedup).
This reduction in timings is due to several contributing factors.
First, the evaluation of the $\hat{T}_1$ contractions in NIVO is sped up by a factor of $\mathcal{O}(n N / N_\mathbf{H})$, where $n$ is the number of commutators included in the BCH series.
Second, the contributions due to doubles, $\hat{O}^{(k+1)}\leftarrow[\hat{C}^{(k)}, \hat{T}_2]$, have identical scaling for the first commutator, but for $k \geq 1$ they can be evaluated with a speedup of a factor of $\mathcal{O}(N^2 / N_\mathbf{H}^2)$.
Third, the NIVO approximation also reduces $t_\mathrm{misc}$ significantly because the tensors transpose and accumulation operations costs are reduced from ${\cal O}(N^4)$ to $\mathcal{O}(N^2N_\mathbf{H}^2)$.
For large $N / N_\mathbf{H}$ ratios, the cost to evaluate $\bar{H}$ in the NIVO approximation is dominated by the commutator $[\hat{H},\hat{T}_2]$, with scaling identical to that of CCSD.
For comparison, the similarity-transformed Hamiltonian can be evaluated in 24 s with \PSI's CCSD, in 121 s with our NIVO-MR-LDSRG(2) code, and 2208 s with the original MR-LDSRG(2) code (in both cases employing an unrestricted implementation and C$_1$ symmetry).

In general, we observe an increase in $t_2$ due to the extra cost to build two-body intermediates from the auxiliary tensors for methods combined with DF.
The traditional and sequential approaches using the DF/NIVO approximations have similar costs, with the latter being slightly faster due to the efficient transformation of the auxiliary tensors [$\tilde{B}$, Eq.~\eqref{eq:B_trans}] afforded by the DF approximation.
For this example, the DF-sq-MR-LDSRG(2)+NIVO computation ran 12 times faster than the one using the original approach.
As we will demonstrate in the next section, this method is as accurate as the MR-LDSRG(2) and, therefore, the method we recommend for large-scale computations.

\section{Results and Discussion}
\label{sec:result}

\subsection{First row diatomic molecules}
\label{subsec:diatomic}

We first benchmark the effect of DF and the NIVO approximation on the traditional and sequential versions of the MR-LDSRG(2).
Our test set consists of eight diatomic molecules: BH, HF, LiF, BeO, CO, \ce{C2}, \ce{N2}, and \ce{F2}.
Specifically, we computed equilibrium bond lengths ($r_e$), harmonic vibrational frequencies ($\omega_e$), anharmonicity constants ($\omega_e x_e$), and dissociation energies ($D_0$) and compare those to experimental data taken from Ref.~\citenum{Huber:1979cc}.
The dissociation energy $D_0$ includes zero-point vibrational energy corrections that account for anharmonicity effects and is computed as $D_0 = D_e - \omega_e /2 + \omega_e x_e /4$ (in a.u.), where $D_e$ is the dissociation energy with respect to the bottom of the potential.\cite{Irikura:2007jg}
Since our current implementation of the MR-DSRG cannot handle half-integer spin states, the energies of the atoms Li, B, C, N, O, and F were computed as half of the energy of the stretched homonuclear diatomic molecule at a distance of 15 \AA{}.
All spectroscopic constants were obtained via a polynomial fit of the energy using nine equally spaced points centered around the equilibrium bond length and separated by 0.2 \AA{}.
For all eight molecules, we adopted a full-valence active space where the 1s orbital of hydrogen, and the 2s and 2p orbitals of first-row elements are considered as active orbitals.
No orbitals were frozen in the CASSCF optimization procedure.
The flow parameter for all DSRG computations was set to $s=0.5$ \sunit, as suggested by our previous work.\cite{Li:2015iz}
All computations utilized the cc-pVQZ basis set\cite{Dunning:1989bx,*Woon:1994jq} and 1s-like orbitals of the first-row elements were frozen in the CC and MR-DSRG treatments of electron correlation.
In DF computations, we employed a mixed flavor of the auxiliary basis sets.
For CASSCF, the cc-pVQZ-JKFIT auxiliary basis set\cite{Weigend:2002ic} was used for H, B, C, N, O and F atoms, and the def2-QZVPP-JKFIT basis set\cite{Weigend:2005gx,*Weigend:2008df} was used for Li and Be atoms.
The cc-pVQZ-RI basis set\cite{Weigend:2002jp,*Hattig:2005dm} was applied to all atoms in DSRG computations.

\begin{figure}[!h]
\ifpreprint
    \includegraphics[width=0.5\columnwidth]{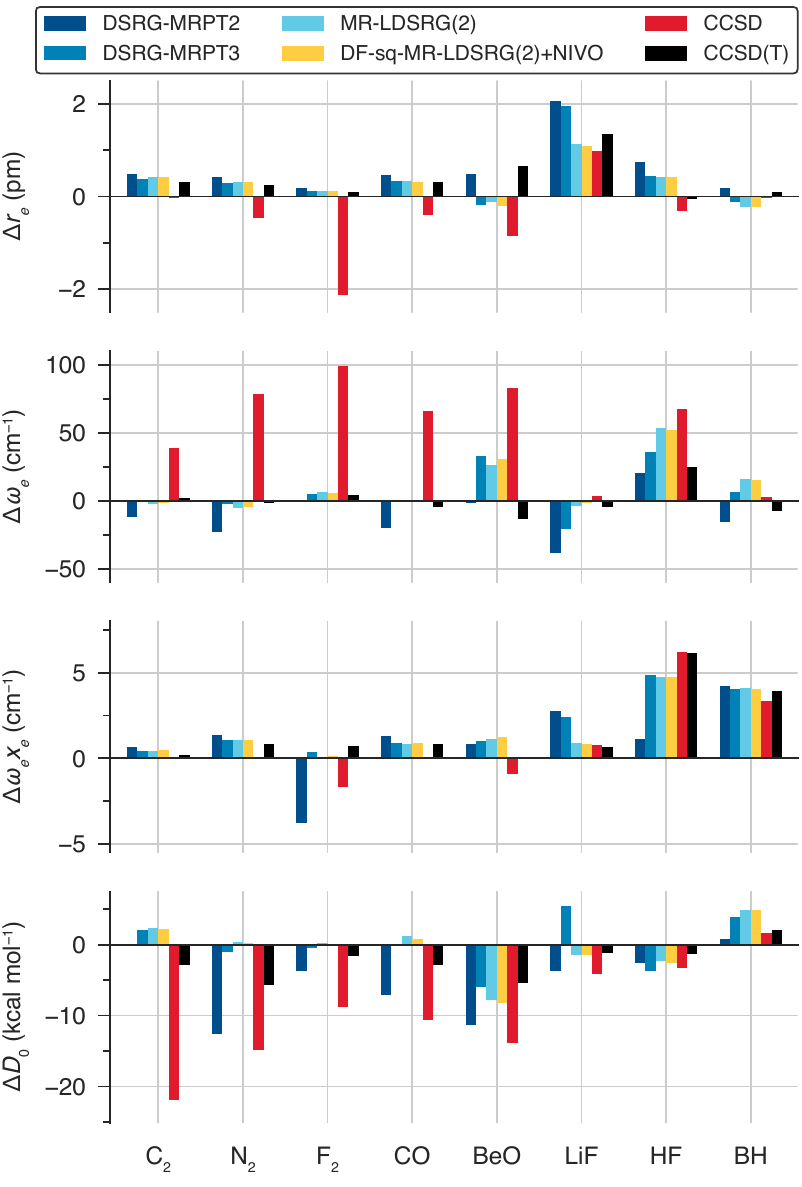}
\else
    \includegraphics[width=0.95\columnwidth]{relaxed_methods_2.pdf}
\fi
    \caption{Comparison of second- and third-order MR-DSRG perturbation theory (DSRG-MRPT2, DSRG-MRPT3), MR-LDSRG(2), DF-sq-MR-LDSRG(2)+NIVO, and single reference coupled cluster methods on a test set composed of 8 diatomic molecules.	Deviations of equilibrium bond lengths ($r_e$), harmonic vibrational frequencies ($\omega_e$), anharmonicity constants ($\omega_e x_e$), and dissociation energies ($D_0$) with respect to experimental values.\cite{Huber:1979cc} All results were computed with cc-pVQZ basis, and core orbitals are frozen in MR-DSRG and coupled cluster computations.}
	\label{fig:rel_methods}
\end{figure}

\newcolumntype{d}[1]{D{.}{.}{#1}}

\begingroup
\begin{table*}[t!]
\begin{threeparttable}

\tiny
\newcommand\separator{4ex}

\ifpreprint
\renewcommand{\arraystretch}{0.85}
\else
\renewcommand{\arraystretch}{1.0}
\fi

\caption{Error statistics for the equilibrium bond lengths ($r_e$, in pm), harmonic vibrational frequencies ($\omega_e$, in \cm), anharmonicity constants ($\omega_e x_e$, in \cm), and dissociation energies ($D_0$, in \kcal) of the eight diatomic molecules computed using various MR-DSRG schemes.
All results were obtained using the cc-pVQZ basis and core orbitals were frozen in the MR-DSRG and CC computations.
The statistical indices are: mean signed error (Mean), mean absolute error (MAE), standard deviation (SD), and maximum absolute error (Max).}

\label{tab:relaxed_summary}

\begin{tabular*}{\textwidth}{@{\extracolsep{\fill} } l l d{3.2} *{12}{d{2.2}} @{}}
       
\hline
\hline

& & \multicolumn{2}{c}{DSRG-MRPT} & \multicolumn{3}{c}{MR-LDSRG(2)} & \multicolumn{3}{c}{sq-MR-LDSRG(2)} & \multicolumn{3}{c}{DF-sq-MR-LDSRG(2)+NIVO} & \multicolumn{1}{c}{\multirow{2}{*}{CCSD}} & \multicolumn{1}{c}{\multirow{2}{*}{CCSD(T)}} \\
\cline{3-4} \cline{5-7} \cline{8-10} \cline{11-13}
& & \multicolumn{1}{c}{PT2} & \multicolumn{1}{c}{PT3} & \multicolumn{1}{c}{Conv.\tnote{a}} & \multicolumn{1}{c}{DF} & \multicolumn{1}{c}{DF+NIVO} & \multicolumn{1}{c}{Conv.\tnote{a}} & \multicolumn{1}{c}{DF} & \multicolumn{1}{c}{DF+NIVO} & \multicolumn{1}{c}{comm(2)} & \multicolumn{1}{c}{comm(3)} & \multicolumn{1}{c}{comm(4)} & &  \\
\hline

\rule{0pt}{0ex}
$r_e$ & Mean & 0.63 & 0.41 & 0.30 & 0.30 & 0.30 & 0.28 & 0.28 & 0.28 & 0.28 & 0.28 & 0.28 & -0.41 & 0.38 \\
& MAE & 0.63 & 0.48 & 0.39 & 0.39 & 0.38 & 0.39 & 0.39 & 0.39 & 0.40 & 0.38 & 0.39 & 0.65 & 0.40 \\
& SD & 0.85 & 0.74 & 0.49 & 0.50 & 0.49 & 0.48 & 0.48 & 0.48 & 0.49 & 0.48 & 0.48 & 0.92 & 0.57 \\
& Max & 2.07 & 1.96 & 1.15 & 1.15 & 1.15 & 1.10 & 1.10 & 1.10 & 1.11 & 1.10 & 1.09 & 2.13 & 1.36 \\

\rule{0pt}{\separator}
$\omega_e$ & Mean & -11.2 & 7.2 & 11.2 & 11.1 & 11.1 & 11.8 & 11.9 & 11.8 & 13.0 & 11.9 & 11.9 & 54.9 & -0.1 \\
& MAE & 16.5 & 13.1 & 14.3 & 14.2 & 14.3 & 13.8 & 13.9 & 13.9 & 16.0 & 13.8 & 13.9 & 54.9 & 7.9 \\
& SD & 20.0 & 19.0 & 22.0 & 21.9 & 22.0 & 21.9 & 22.1 & 22.1 & 26.3 & 21.6 & 22.2 & 64.5 & 10.7 \\
& Max & 38.3 & 36.1 & 53.2 & 52.8 & 53.7 & 51.2 & 51.8 & 51.9 & 64.5 & 50.3 & 52.3 & 99.3 & 24.5 \\

\rule{0pt}{\separator}
$\omega_e x_e$ & Mean & 1.1 & 1.9 & 1.7 & 1.5 & 1.7 & 1.7 & 1.8 & 1.7 & 1.9 & 1.6 & 1.7 & 1.0 & 1.7 \\
& MAE & 2.0 & 1.9 & 1.7 & 1.5 & 1.7 & 1.7 & 1.8 & 1.7 & 1.9 & 1.6 & 1.7 & 1.6 & 1.7 \\
& SD & 2.4 & 2.5 & 2.3 & 2.1 & 2.3 & 2.3 & 2.6 & 2.3 & 2.6 & 2.2 & 2.3 & 2.6 & 2.6 \\
& Max & 4.2 & 4.9 & 4.7 & 4.1 & 4.8 & 4.8 & 5.7 & 4.7 & 5.9 & 4.4 & 4.8 & 6.2 & 6.2 \\

\rule{0pt}{\separator}
$D_0$ & Mean & -5.0 & 0.1 & -0.3 & -0.1 & -0.0 & -0.6 & -0.3 & -0.5 & -1.0 & -0.5 & -0.3 & -9.5 & -2.4 \\
& MAE & 5.3 & 2.8 & 2.6 & 2.8 & 2.6 & 2.6 & 2.9 & 2.5 & 3.1 & 2.7 & 2.8 & 9.9 & 2.9 \\
& SD & 6.8 & 3.5 & 3.5 & 3.7 & 3.5 & 3.7 & 3.8 & 3.6 & 4.0 & 3.7 & 3.8 & 11.8 & 3.3 \\
& Max & 12.7 & 6.0 & 7.9 & 7.8 & 7.5 & 8.4 & 8.4 & 8.3 & 8.6 & 8.4 & 8.2 & 21.9 & 5.6 \\

\hline
\hline
\end{tabular*}

\begin{tablenotes}
\item[a] Computed using conventional four-index two-electron integrals.
\end{tablenotes}

\end{threeparttable}
\end{table*}
\endgroup

Figure~\ref{fig:rel_methods} and Table~\ref{tab:relaxed_summary} report a comparison of second- and third-order DSRG multireference perturbation theory (DSRG-MRPT2/3), the original MR-LDSRG(2), DF-sq-LDSRG(2)+NIVO, CCSD, and CCSD(T). 
The mean absolute error (MAE) and standard deviation (SD) reported in Table~\ref{tab:relaxed_summary} show that MR-LDSRG(2) method is as accurate as CCSD(T) in predicting $r_e$, $\omega_e x_e$ and $D_0$, while it predicts $\omega_e$ that are of accuracy between that of CCSD and CCSD(T).

The fact that the MR-LDSRG(2) results are more accurate than those from CCSD suggests that the full-valence treatment of static correlation leads to a more accurate treatment of correlation. It is also rewarding to see that in many instances the MR-LDSRG(2) has an accuracy similar to that of CCSD(T), despite the fact that the former does not include triples corrections.

To analyze the impact of each approximation of the MR-LDSRG(2) method, in Table~\ref{tab:relaxed_rel_error} we report the mean absolute difference between properties computed with and without each approximation.
The use of a sequential ansatz has a modest effect on all properties, with the largest mean absolute differences observed for $\omega_e$ (1.6 \cm) and $D_0$ (0.3 \kcal). Nevertheless, the MAE with respect experimental results is nearly unchanged, if not slightly improving.
The DF and NIVO approximations have an effect on molecular properties that is comparable in magnitude and smaller than the deviation introduced by the sequential ansatz.
When these three approximations are combined together, the resulting method shows errors with respect to experimental values that are nearly identical to those from the conventional MR-LDSRG(2).
The only noticeable deviations are found for $\omega_e$ (MAE 13.9 vs.~14.3 \cm) and $D_0$ (MAE 2.6 vs.~2.5 \kcal).

\begin{table*}[t!]
\begin{threeparttable}

\newcommand\separator{3ex}
\footnotesize

\ifpreprint
\renewcommand{\arraystretch}{0.85}
\else
\renewcommand{\arraystretch}{1.0}
\fi

\caption{The mean absolute differences in predicting equilibrium bond lengths ($r_e$, in pm), harmonic vibrational frequencies ($\omega_e$, in \cm), anharmonicity constants ($\omega_e x_e$, in \cm), and dissociation energies ($D_0$, in \kcal) of the eight diatomic molecules between method pairs that differ by only one technique introduced in this report.
  Techniques include: sequential transformation (ST), density fitting (DF), NIVO approximation, and commutator truncation of the BCH expansion [comm($k$), $k = 2,3,4$].
  All results were computed using the cc-pVQZ basis set and core orbitals were frozen in the MR-DSRG computations.}

\label{tab:relaxed_rel_error}

\begin{tabular*}{\textwidth}{@{\extracolsep{\fill} } *{3}{l} *{4}{c} @{}}
       
\hline
\hline

\multicolumn{1}{c}{\multirow{2}{*}{Technique}} & \multicolumn{2}{c}{\multirow{2}{*}{MR-LDSRG(2) method pair}} & \multicolumn{4}{c}{Mean absolute difference} \\
\cline{4-7}
& & & \multicolumn{1}{c}{$r_e$} & \multicolumn{1}{c}{$\omega_e$} & \multicolumn{1}{c}{$\omega_e x_e$} & \multicolumn{1}{c}{$D_0$} \\
\hline

\multirow{2}{*}{ST} & Original & ST & 0.03   & 1.6   & 0.0   & 0.3   \\
& DF & DF/ST & 0.03   & 1.5   & 0.3   & 0.2   \\

\hline
\multirow{2}{*}{DF} & Original & DF & 0.00   & 0.2   &  0.1   &   0.2   \\
& ST & DF/ST &  0.00   &   0.2   &   0.1   &   0.3   \\

\hline
\multirow{2}{*}{NIVO} & DF & DF/NIVO & 0.01   &  0.4   &   0.1   &   0.2   \\
& DF/ST & DF/ST/NIVO & 0.01   &  0.3   &  0.1   &  0.3   \\

\hline
comm(4) & DF/ST/NIVO & DF/ST/NIVO/comm(4) & 0.00   &   0.1   &   0.0   &   0.3   \\

\hline
4$^{\rm th}$ comm. & DF/ST/NIVO/comm(4) & DF/ST/NIVO/comm(3) & 0.01   &   0.8   &  0.1   &  0.4   \\

\hline
3$^{\rm rd}$ comm. & DF/ST/NIVO/comm(3) & DF/ST/NIVO/comm(2) & 0.04   &   4.3   &   0.3   &  0.6   \\

\hline
\hline
\end{tabular*}
\end{threeparttable}
\end{table*}

In this study, we also investigate the effect of combining the DF-sq-LDSRG(2)+NIVO method with truncation of the BCH expansion, i.e., terminating $\bar{H}(s) = \tilde{H} (s) + \sum_{n=1}^{k} \frac{1}{n!} \hat{C}^{(n)} (s)$ at a given integer $k$.
The recursive evaluation of $\bar{H}(s)$ via Eq.~\eqref{eq:recursive} usually requires 10--12.
Truncation of the BCH series to a few terms may therefore introduce speedups of up to 3--4 times.
In Table~\ref{tab:relaxed_summary} and Table~\ref{tab:relaxed_rel_error} we report statistics computed by approximating the BCH expansion up to  2, 3, and 4 commutators.
The use of only two commutator introduces noticeable deviations with respect to experiments for $\omega_e$ and $D_0$. Compared to the full BCH series, this truncation level increases the MAE of $\omega_e$ and $D_0$ by 12.6 \cm and  $0.5$ \kcal, respectively.
The inclusion of the triply-nested commutator significantly reduces these deviations to only 0.1 \cm and 0.2  \kcal, respectively.
The four-fold commutator term yields $r_e$, $\omega_e$, and $\omega_e x_e$ that are nearly identical to those from the untruncated BCH series, while the MAE of $D_0$ deviates only by 0.3 \kcal.

Since the error introduced by neglecting the four commutator term is smaller or comparable to the other approximations considered here, our results suggest that a BCH series truncated to three commutators may offer a good compromise between accuracy and speed.

\subsection{Cyclobutadiene}
\label{subsec:cyclobutadiene}

Next, we consider the automerization reaction of cyclobutadiene (CBD, C$_4$H$_4$). We study the energy difference between the rectangular ($D_{\rm 2h}$) energy minimum and the square transition state ($D_{\rm 4h}$).\cite{Shen:2012kn}
This reaction is a challenging chemistry problem for both experiment and theory.\cite{Whitman:1982hy,Nakamura:1989ff, Balkova:1994kp, SanchoGarcia:2000dx, Levchenko:2003hq, EckertMaksic:2006ex,Shen:2012kn,Lyakh:2011iq,Shen:2008kv,Li:2009eq,BhaskaranNair:2008dj,Demel:2008ey,Whitman:1982hy,EckertMaksic:2006ex, Wu:2012ek}
Due to its instability, there are no direct measurements of the reaction barrier, and  experiments performed on substituted cyclobutadienes  suggest the barrier height falls in the range 1.6--10 \kcal.\cite{Whitman:1982hy}
In this work, we optimized the equilibrium and transition state geometries using finite differences of energies to compute the barrier height.
Specifically, we compare both DF-MR-LDSRG(2)+NIVO and DF-sq-MR-LDSRG(2)+NIVO optimized geometries to those obtained from the state-specific MRCC of Mukherjee and co-workers (Mk-MRCC)\cite{Mahapatra:1998cp,Mahapatra:1999bp,*Chattopadhyay:2000io,*Chattopadhyay:2004fw,*Pahari:2004ih} as implemented in \PSI.
\cite{Evangelista:2006gf,*Evangelista:2008gv,*Evangelista:2010cq}

To reduce computational cost, all MR-DSRG calculations performed two steps of the reference relaxation procedure discussed in Sec.~\ref{subsec:mrdsrg}.
A comparison of this procedure with full reference relaxation using the cc-pVDZ basis set shows errors of ca. 0.01 \kcal for absolute energies, 0.0001 \AA{} for bond lengths, and 0.001$^{\circ}$ for bond angles.
We applied a Tikhonov regularization denominator shift\cite{Taube:2009jz} of 1 m\Eh in all Mk-MRCC calculations to guarantee convergence.
The Mk-MRCC implementation used in this work neglects effective Hamiltonian couplings between reference determinants that differ by three or more spin orbitals, and therefore yield approximate results when applied to the CAS(4e,4o) reference considered here.
All computations utilized the cc-pV$X$Z ($X$=D, T, Q, 5) basis set\cite{Dunning:1989bx,*Woon:1994jq}, and the corresponding cc-pV$X$Z-JKFIT\cite{Weigend:2002ic} and cc-pV$X$Z-RI\cite{Weigend:2002jp,*Hattig:2005dm} auxiliary basis sets for DF-CASSCF and DF-DSRG computations, respectively.
The 1s core electrons of carbon atoms were frozen in all post-CASSCF methods.
All results were computed using semi-canonical CASSCF orbitals.

\begin{figure}[!hp]
\ifpreprint
    \includegraphics[width=0.6\columnwidth]{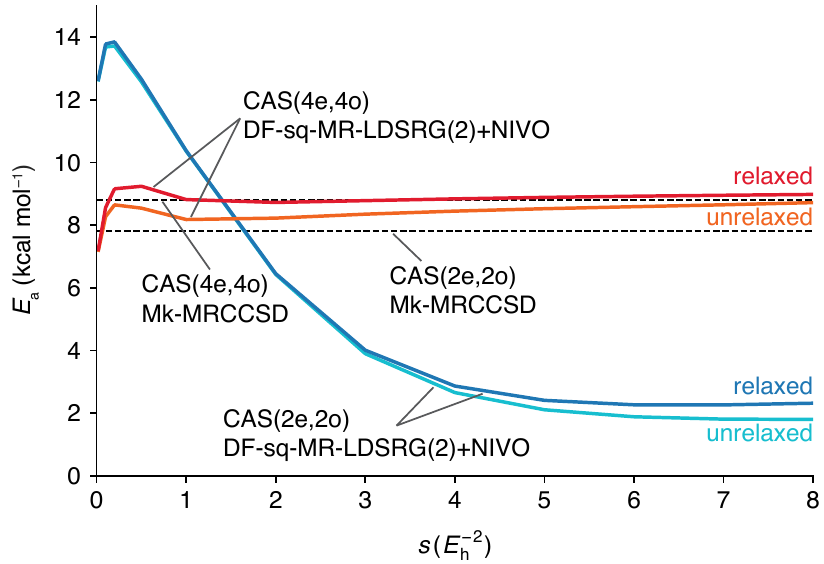}
\else
    \includegraphics[width=0.9\columnwidth]{C4H4s-t_DZ_gap_ai.pdf}
\fi
\caption{The automerization barrier ($E_\text{a}$) of cyclobutadiene computed using the DF-sq-MR-LDSRG(2)+NIVO theory with varying flow parameters. Results were obtained using the cc-pVDZ basis set. We also applied a Tikhonov regularization denominator shift\cite{Taube:2009jz} of 1 m\Eh in all Mk-MRCC calculations to guarantee convergence.}
\label{fig:C4H4s-t_DZ_gap}
\end{figure}

Preliminary computations using the cc-pVDZ basis using the CAS(2e,2o) and CAS(4e,4o) active spaces revealed an interesting aspect of this system.
As shown in Fig.~\ref{fig:C4H4s-t_DZ_gap}, the $s$-dependency of the automerization barrier displays significantly different behavior for these two active spaces.
In both cases, the predicted activation energies change significantly for small values of $s$ ($<0.2$ \sunit), a normal trend observed for all DSRG computations and due to the increased recovery of dynamical correlation energy.
Interestingly, while the CAS(4e,4o) curve flattens out for larger values of $s$, the CAS(2e,2o) curve shows a significant $s$-dependence in the range $s \in [0.5,8]$ \sunit.

\begin{figure}[!htp]
\ifpreprint
    \includegraphics[width=0.7\columnwidth]{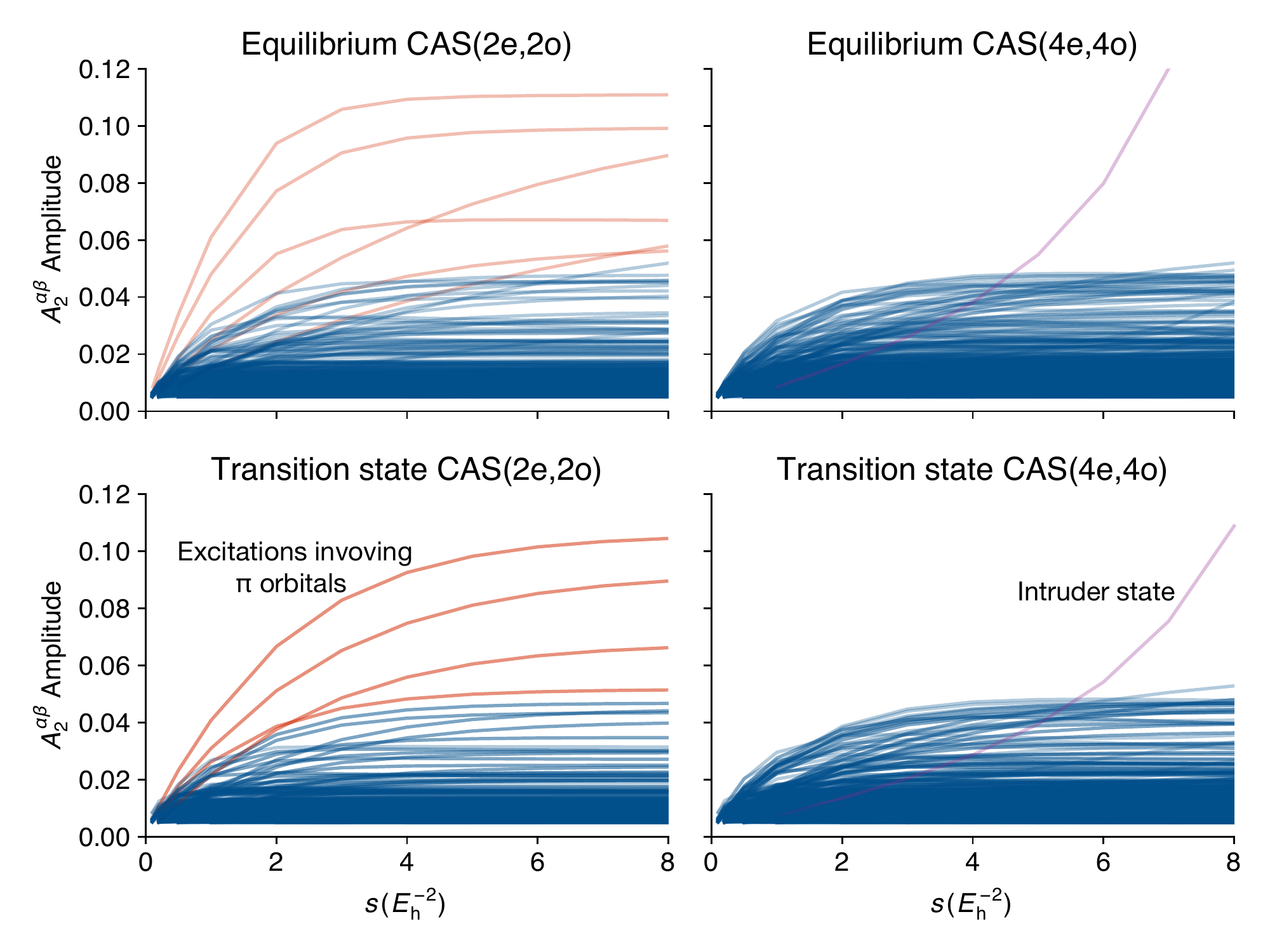}
\else
    \includegraphics[width=0.95\columnwidth]{T2ABplotintru.pdf}
\fi
\caption{The unrelaxed DF-sq-MR-LDSRG(2)+NIVO/cc-pVDZ double substitution amplitudes involving both alpha and beta electrons ($A_2^{\alpha\beta}$) as a function of the flow parameter $s$ for the rectangular equilibrium and the square transition state of cyclobutadiene.}
\label{fig:T2ABplotintru}
\end{figure}

To understand the origin of this difference we analyze the double substitution amplitudes [$\tens{t}{ab}{ij}(s)$] as a function of $s$ for both the equilibrium and transition state geometries, as shown in Fig.~\ref{fig:T2ABplotintru}.
In the CAS(2e,2o) case, we notice some abnormally large amplitudes (indicated in red), some of which are as large as 0.1. These amplitudes correspond to excitations within the four $\pi$ orbitals of CBD, and suggests that the CAS(2e,2o) space is insufficient to capture all static correlation effects in CBD.
The offending amplitudes converge at different rates as $s$ increases, and introduce a strong $s$-dependence in the energy barrier. Note also that in the limit of $s \rightarrow \infty$ there is a significant difference in the barrier for the CAS(2e,2o) and CAS(4e,4o) spaces.

In contrast, in the CAS(4e,4o) computations all excitations within the $\pi$ orbitals are included in the active space and the resulting DSRG amplitudes have absolute values less than 0.05.
Diverging amplitudes in computations with CAS(4e,4o) reference wave functions, corresponding to intruder states, can also be seen in Fig.~\ref{fig:T2ABplotintru}.
Our results reported in Table~\ref{tab:C4H4geoms} are all based on the flow parameter value $s=1.0$ \sunit, which is significantly far from the region  ($s > 5.0$ \sunit) where amplitudes begin to diverge, and at the same time leads to well converged absolute energies (see Figure S3 in the Supporting Information).
We also performed computations using $s$ = 0.5 \sunit (reported in Table S4) to verify that the automerization energies computed with different values of the flow parameter are consistent.
In general, the difference in automerization energies computed with $s = 0.5$ and 1.0 \sunit is of the order of 0.6--0.7 \kcal.
Note that intruder states are also encountered in Mk-MRCCSD computations based on the CAS(4e,4o) reference, and lead to convergence issues that could be avoided only via Tikhonov regularization.

\begin{table*}[t!]
\begin{threeparttable}
\footnotesize
\caption{Automerization reaction barrier ($E_{\rm a}$, in \kcal) and geometry parameters (bond lengths in \AA, bond angles in degree) of cyclobutadiene. All DSRG computations used $s = 1.0$ \sunit. All computations employed the CAS(4e,4o) reference and core orbitals constructed from carbon 1s orbitals were frozen for MR-DSRG and MRCC computations. As such, $n_{\bf C} = 8$ and $n_{\bf A} = 4$, where $n_{\bf X} = N_{\bf X} / 2$ ($\bf X \in \{C, A, V, H, P, G\}$) is the number molecular spacial orbitals for space $\bf X$.}
\label{tab:C4H4geoms}

\ifpreprint
\renewcommand{\arraystretch}{0.85}
\else
\renewcommand{\arraystretch}{0.95}
\fi

\begin{tabular*}{\textwidth}{@{\extracolsep{\fill} } ld{2.2}d{1.4}d{1.4}d{1.4}d{3.2}d{1.4}d{1.4} @{}}
\hline
\hline

\multicolumn{1}{c}{\multirow{2}{*}{Method}} & \multicolumn{1}{c}{\multirow{2}{*}{$E_\text{a}$}} & \multicolumn{4}{c}{D$_{\rm 2h}$} & \multicolumn{2}{c}{D$_{\rm 4h}$} \\
\cline{3-6} \cline{7-8}
& & \multicolumn{1}{c}{\ce{C-C}\tnote{a}} &  \multicolumn{1}{c}{\ce{C=C}\tnote{a}} & \multicolumn{1}{c}{\ce{C-H}} &  \multicolumn{1}{c}{\ce{$\angle$ C-C-H}\tnote{b}} & \multicolumn{1}{c}{\ce{C-C}} & \multicolumn{1}{c}{\ce{C-H}} \\
\hline

\noalign{\vskip 2pt}
\multicolumn{8}{c}{cc-pVDZ ($n_{\bf V} = 60$)} \\
\hline
              CASSCF &  6.49  &  1.5502  &  1.3567  &  1.0790  &  134.87  &  1.4472  &  1.0779  \\
    DF-MR-LDSRG(2)+NIVO &  8.56 &  1.5769 &  1.3660 &  1.0945 &  134.92 &  1.4624 &  1.0932 \\
DF-sq-MR-LDSRG(2)+NIVO &  8.62  &  1.5768  &  1.3659  &  1.0944  &  134.92  &  1.4623  &  1.0931  \\
           Mk-MRCCSD &  8.80  &  1.5733  &  1.3623  &  1.0931  &  134.91  &  1.4585  &  1.0920  \\
        Mk-MRCCSD(T) &  7.56  &  1.5772  &  1.3699  &  1.0951  &  134.92  &  1.4652  &  1.0941  \\

\noalign{\vskip 2pt}
\multicolumn{8}{c}{cc-pVTZ ($n_{\bf V}=160$)} \\
\hline
              CASSCF &  7.44  &  1.5475  &  1.3471  &  1.0694  &  134.83  &  1.4409  &  1.0683  \\
    DF-MR-LDSRG(2)+NIVO &  9.87 &  1.5668 &  1.3488 &  1.0789 &  134.91 &  1.4483 &  1.0775 \\
DF-sq-MR-LDSRG(2)+NIVO &  9.93  &  1.5666  &  1.3487  &  1.0788  &  134.91  &  1.4481  &  1.0774  \\
           Mk-MRCCSD & 10.09  &  1.5628  &  1.3452  &  1.0775  &  134.89  &  1.4442  &  1.0764  \\
        Mk-MRCCSD(T) &  8.56  &  1.5671  &  1.3535  &  1.0797  &  134.90  &  1.4515  &  1.0786  \\

\noalign{\vskip 2pt}
\multicolumn{8}{c}{cc-pVQZ ($n_{\bf V}=324$)} \\
\hline
              CASSCF &  7.53  &  1.5467  &  1.3462  &  1.0689  &  134.84  &  1.4400  &  1.0678  \\
    DF-MR-LDSRG(2)+NIVO & 10.16 &  1.5634 &  1.3452 &  1.0782 &  134.96 &  1.4447 &  1.0768 \\
DF-sq-MR-LDSRG(2)+NIVO & 10.21  &  1.5631  &  1.3451  &  1.0781  &  134.96  &  1.4446  &  1.0766  \\
           Mk-MRCCSD & 10.28  &  1.5591  &  1.3417  &  1.0768  &  134.94  &  1.4406  &  1.0756  \\
        Mk-MRCCSD(T) &  8.69  &  1.5634  &  1.3500  &  1.0791  &  134.95  &  1.4480  &  1.0779  \\

\noalign{\vskip 2pt}
\multicolumn{8}{c}{cc-pV5Z ($n_{\bf V}=568$)} \\
\hline
    DF-MR-LDSRG(2)+NIVO\tnote{c} & 10.26 \\
DF-sq-MR-LDSRG(2)+NIVO\tnote{c} & 10.30 \\
                  
\hline
\hline
\end{tabular*}

\begin{tablenotes}
\item[a] \ce{C-C} and \ce{C=C} refer to the longer and shorter carbon--carbon bonds, respectively.
\item[b] \ce{$\angle$ C-C-H} is the bond angle between the \ce{C-H} bond and the longer \ce{C-C} bond.
\item[c] Based on the corresponding cc-pVQZ optimized geometries.
\end{tablenotes}

\end{threeparttable}
\end{table*}

Geometric parameters for the optimized structures and energy barriers of CBD computed with the CAS(4e,4o) reference are reported in Table~\ref{tab:C4H4geoms}.
A comparison the energy barrier computed at the CASSCF and correlated levels shows that dynamical correlation is important in this system as it increases is by about 1--3 \kcal.
Our best estimate for the automerization barrier of CBD is 10.3 \kcal at the DF-sq-MR-LDSRG(2)+NIVO/cc-pV5Z level of theory.
This value is likely to be slightly higher then the exact results since in the Mk-MRCC results perturbative triples corrections contribute to lowering the barrier by ca. 1.5 \kcal.
In general, the MR-LDSRG(2) results are between those of Mk-MRCCSD and Mk-MRCCSD(T), reinforcing the same observation we made in the benchmark of diatomic molecules.
For instance, the MR-LDSRG(2) predicted \ce{C-C} bond length at the $D_{\rm 4h}$ geometry is 1.4446 \AA{}, which is almost midway between the Mk-MRCCSD (1.4406 \AA) and Mk-MRCCSD(T) (1.4480 \AA) values.
As expected, the differences between the conventional and sequentially transformed MR-DSRG(2) results are negligible.
Our results in cc-pVDZ and cc-pVTZ bases also agree well with other reported theoretical values, especially those computed with multireference methods,\cite{Shen:2012kn,Lyakh:2011iq,Shen:2008kv,Li:2009eq,Balkova:1994kp,BhaskaranNair:2008dj,EckertMaksic:2006ex,Demel:2008ey,EckertMaksic:2006ex} and the experimental range reported in Ref.~\citenum{Whitman:1982hy}.
Using our new implementation, we were able to perform, for the first time, nonperturbative multireference computations on cyclobutadiene using the cc-pV5Z basis (580 correlated orbitals) on a single node with 128 GB of memory.

\section{Conclusion}
\label{sec:conclusion}

In this work, we describe a strategy to reduce the computational and memory costs of the multireference driven similarity renormalization group (MR-DSRG). We demonstrate that the cost of the linear MR-DSRG with singles and doubles [MR-LDSRG(2)] can be lowered substantially without compromising its accuracy by using a combination of: 1) a sequential unitary transformation, 2) density fitting (DF) of the  two-electron integrals, and 3) the non-interacting virtual orbital (NIVO) operator approximation.
The sequential MR-DSRG scheme introduced in this work [sq-MR-DSRG] reduces the cost of evaluating single-excitations and allows to treat them exactly. Like in the case of Brueckner coupled cluster theory,\cite{Chiles:1981jg, Handy:1989fn, Stanton:1998fk} this approach reduces the number of algebraic terms in the DSRG equations because there are no terms (diagrams) containing single excitations.
The use of DF integrals reduces the memory requirements of the original MR-DSRG(2) from $\mathcal{O}(N^4)$ to ${\cal O}(N^2M)$, where $N$ is the number of basis functions.
Density fitting is particularly convenient when combined with the sequential approach because the contributions of singles can be directly included in the DF auxiliary three-index integrals, reducing the integral transformation cost from ${\cal O} (N^5)$ to ${\cal O} (N^3 M)$, where $M$ is the number of auxiliary basis functions.
The NIVO approximation neglects the operator components of a commutator with three or more virtual indices.
A formal analysis of this approximation showed that the leading error is of fourth order in perturbation theory.
In practice, NIVO is crucial to both avoiding the memory bottleneck of the MR-DSRG(2) and reducing the computational cost to evaluate the transformed Hamiltonian.

To benchmark the MR-LDSRG(2) and sq-MR-LDSRG(2) approaches and assess the impact of the DF and  NIVO approximations, we computed the spectroscopic constants of eight diatomic molecules using the full-valence active space and the cc-pVQZ basis set.
Compared to experimental data, both MR-DSRG methods yield results that are as accurate as those obtained with CCSD(T).
Moreover, the DF-sq-MR-LDSRG(2)+NIVO results are almost identical to those computed without the NIVO approximation: the harmonic vibrational frequencies, anharmonicity constants, and dissociation energies only differ by, on average, 0.1 \cm, 0.1 \cm, and 0.2 \kcal, respectively.
These results supports our claim that the speedup brought by the NIVO approximation does not sacrifice the accuracy of both variants of the MR-LDSRG(2).

Combining DF and the NIVO approximation, both the traditional and sequential MR-LDSRG(2) can be routinely applied to chemical systems with more than 500 basis function.
We demonstrate this point by studying the automerization reaction of cyclobutadiene using a quintuple-$\zeta$ basis set (584 basis functions).
Our best estimate of the reaction barrier from DF-sq-MR-LDSRG(2)+NIVO/cc-pV5Z is 10.3 \kcal. However, we expect that this result is likely overestimated due to the lack of three-body corrections in the MR-LDSRG(2) theory.
Our results agree well with Mk-MRCCSD predictions and multireference coupled cluster reported in the literature.

In conclusion, we have shown that it is possible to significantly reduce the cost of MR-LDSRG(2) computations without reducing the accuracy of this approach. The sequential approach and NIVO approximations are general, and can be applied to improve the efficiency of other unitary nonperturbative methods (e.g., unitary coupled cluster theory) and downfolding schemes for classical-quantum hybrid algorithms.\cite{Bauman:2019wj}
\section*{Supporting Information}
%
Detailed timings and formal scaling required to evaluate the transformed Hamiltonian, error statistics for the eight diatomic molecules, cyclobutadiene automerization energies reported in the literature, and convergence of DF-sq-MR-LDSRG(2)+NIVO absolute energy with respect to flow parameter.

%
%
\section*{Acknowledgments}
This work was supported by the U.S. Department of Energy under Award No. DE-SC0016004, a Research Fellowship of the Alfred P. Sloan Foundation, and a Camille Dreyfus Teacher-Scholar Award.

\footnotesize
\mciteErrorOnUnknownfalse
\bibliography{ms_seq_df_dsrg_ext,ms_seq_df_dsrg}

\providecommand{\latin}[1]{#1}
\makeatletter
\providecommand{\doi}
  {\begingroup\let\do\@makeother\dospecials
  \catcode`\{=1 \catcode`\}=2\doi@aux}
\providecommand{\doi@aux}[1]{\endgroup\texttt{#1}}
\makeatother
\providecommand*\mcitethebibliography{\thebibliography}
\csname @ifundefined\endcsname{endmcitethebibliography}
  {\let\endmcitethebibliography\endthebibliography}{}
\begin{mcitethebibliography}{155}
\providecommand*\natexlab[1]{#1}
\providecommand*\mciteSetBstSublistMode[1]{}
\providecommand*\mciteSetBstMaxWidthForm[2]{}
\providecommand*\mciteBstWouldAddEndPuncttrue
  {\def\EndOfBibitem{\unskip.}}
\providecommand*\mciteBstWouldAddEndPunctfalse
  {\let\EndOfBibitem\relax}
\providecommand*\mciteSetBstMidEndSepPunct[3]{}
\providecommand*\mciteSetBstSublistLabelBeginEnd[3]{}
\providecommand*\EndOfBibitem{}
\mciteSetBstSublistMode{f}
\mciteSetBstMaxWidthForm{subitem}{(\alph{mcitesubitemcount})}
\mciteSetBstSublistLabelBeginEnd
  {\mcitemaxwidthsubitemform\space}
  {\relax}
  {\relax}

\bibitem[Andersson \latin{et~al.}(1990)Andersson, Malmqvist, Roos, Sadlej, and
  Wolinski]{Andersson:1990jz}
Andersson,~K.; Malmqvist,~P.-{\AA}.; Roos,~B.~O.; Sadlej,~A.~J.; Wolinski,~K.
  {Second-order perturbation theory with a CASSCF reference function}. \emph{J.
  Phys. Chem.} \textbf{1990}, \emph{94}, 5483--5488\relax
\mciteBstWouldAddEndPuncttrue
\mciteSetBstMidEndSepPunct{\mcitedefaultmidpunct}
{\mcitedefaultendpunct}{\mcitedefaultseppunct}\relax
\EndOfBibitem
\bibitem[Andersson \latin{et~al.}(1992)Andersson, Malmqvist, and
  Roos]{Andersson:1992cq}
Andersson,~K.; Malmqvist,~P.-{\AA}.; Roos,~B.~O. {Second-order perturbation
  theory with a complete active space self-consistent field reference
  function}. \emph{J. Chem. Phys.} \textbf{1992}, \emph{96}, 1218--1226\relax
\mciteBstWouldAddEndPuncttrue
\mciteSetBstMidEndSepPunct{\mcitedefaultmidpunct}
{\mcitedefaultendpunct}{\mcitedefaultseppunct}\relax
\EndOfBibitem
\bibitem[Hirao(1992)]{Hirao:1992bq}
Hirao,~K. {Multireference M{\o}ller{\textemdash}Plesset method}. \emph{Chem.
  Phys. Lett.} \textbf{1992}, \emph{190}, 374--380\relax
\mciteBstWouldAddEndPuncttrue
\mciteSetBstMidEndSepPunct{\mcitedefaultmidpunct}
{\mcitedefaultendpunct}{\mcitedefaultseppunct}\relax
\EndOfBibitem
\bibitem[Nakano(1993)]{Nakano:1993hv}
Nakano,~H. {Quasidegenerate perturbation theory with multiconfigurational
  self-consistent-field reference functions}. \emph{J. Chem. Phys.}
  \textbf{1993}, \emph{99}, 7983--7992\relax
\mciteBstWouldAddEndPuncttrue
\mciteSetBstMidEndSepPunct{\mcitedefaultmidpunct}
{\mcitedefaultendpunct}{\mcitedefaultseppunct}\relax
\EndOfBibitem
\bibitem[Werner(1996)]{Werner:1996in}
Werner,~H.-J. {Third-order multireference perturbation theory The CASPT3
  method}. \emph{Mol. Phys.} \textbf{1996}, \emph{89}, 645--661\relax
\mciteBstWouldAddEndPuncttrue
\mciteSetBstMidEndSepPunct{\mcitedefaultmidpunct}
{\mcitedefaultendpunct}{\mcitedefaultseppunct}\relax
\EndOfBibitem
\bibitem[Mahapatra \latin{et~al.}(1999)Mahapatra, Datta, and
  Mukherjee]{SinhaMahapatra:1999bp}
Mahapatra,~U.~S.; Datta,~B.; Mukherjee,~D. {Molecular Applications of a
  Size-Consistent State-Specific Multireference Perturbation Theory with
  Relaxed Model-Space Coefficients}. \emph{J. Phys. Chem. A} \textbf{1999},
  \emph{103}, 1822--1830\relax
\mciteBstWouldAddEndPuncttrue
\mciteSetBstMidEndSepPunct{\mcitedefaultmidpunct}
{\mcitedefaultendpunct}{\mcitedefaultseppunct}\relax
\EndOfBibitem
\bibitem[Angeli \latin{et~al.}(2001)Angeli, Cimiraglia, Evangelisti, Leininger,
  and Malrieu]{Angeli:2001bg}
Angeli,~C.; Cimiraglia,~R.; Evangelisti,~S.; Leininger,~T.; Malrieu,~J.~P.
  {Introduction of \emph{n}-electron valence states for multireference
  perturbation theory}. \emph{J. Chem. Phys.} \textbf{2001}, \emph{114},
  10252--10264\relax
\mciteBstWouldAddEndPuncttrue
\mciteSetBstMidEndSepPunct{\mcitedefaultmidpunct}
{\mcitedefaultendpunct}{\mcitedefaultseppunct}\relax
\EndOfBibitem
\bibitem[Angeli \latin{et~al.}(2007)Angeli, Pastore, and
  Cimiraglia]{Angeli:2007by}
Angeli,~C.; Pastore,~M.; Cimiraglia,~R. {New perspectives in multireference
  perturbation theory: the \emph{n}-electron valence state approach}.
  \emph{Theor. Chem. Acc.} \textbf{2007}, \emph{117}, 743--754\relax
\mciteBstWouldAddEndPuncttrue
\mciteSetBstMidEndSepPunct{\mcitedefaultmidpunct}
{\mcitedefaultendpunct}{\mcitedefaultseppunct}\relax
\EndOfBibitem
\bibitem[Liu(1973)]{Liu:1973fn}
Liu,~B. {\emph{Ab initio} potential energy surface for linear H$_{3}$}.
  \emph{J. Chem. Phys.} \textbf{1973}, \emph{58}, 1925--1937\relax
\mciteBstWouldAddEndPuncttrue
\mciteSetBstMidEndSepPunct{\mcitedefaultmidpunct}
{\mcitedefaultendpunct}{\mcitedefaultseppunct}\relax
\EndOfBibitem
\bibitem[Werner and Knowles(1988)Werner, and Knowles]{Werner:1988ku}
Werner,~H.-J.; Knowles,~P.~J. {An efficient internally contracted
  multiconfiguration{\textendash}reference configuration interaction method}.
  \emph{J. Chem. Phys.} \textbf{1988}, \emph{89}, 5803--5814\relax
\mciteBstWouldAddEndPuncttrue
\mciteSetBstMidEndSepPunct{\mcitedefaultmidpunct}
{\mcitedefaultendpunct}{\mcitedefaultseppunct}\relax
\EndOfBibitem
\bibitem[Knowles and Werner(1988)Knowles, and Werner]{Knowles:1988hv}
Knowles,~P.~J.; Werner,~H.-J. {An efficient method for the evaluation of
  coupling coefficients in configuration interaction calculations}. \emph{Chem.
  Phys. Lett.} \textbf{1988}, \emph{145}, 514--522\relax
\mciteBstWouldAddEndPuncttrue
\mciteSetBstMidEndSepPunct{\mcitedefaultmidpunct}
{\mcitedefaultendpunct}{\mcitedefaultseppunct}\relax
\EndOfBibitem
\bibitem[Hanrath and Engels(1997)Hanrath, and Engels]{Hanrath:1997ch}
Hanrath,~M.; Engels,~B. {New algorithms for an individually selecting MR-CI
  program}. \emph{Chem. Phys.} \textbf{1997}, \emph{225}, 197--202\relax
\mciteBstWouldAddEndPuncttrue
\mciteSetBstMidEndSepPunct{\mcitedefaultmidpunct}
{\mcitedefaultendpunct}{\mcitedefaultseppunct}\relax
\EndOfBibitem
\bibitem[Szalay \latin{et~al.}(2012)Szalay, M{\"u}ller, Gidofalvi, Lischka, and
  Shepard]{Szalay:2012df}
Szalay,~P.~G.; M{\"u}ller,~T.; Gidofalvi,~G.; Lischka,~H.; Shepard,~R.
  {Multiconfiguration Self-Consistent Field and Multireference Configuration
  Interaction Methods and Applications}. \emph{Chem. Rev.} \textbf{2012},
  \emph{112}, 108--181\relax
\mciteBstWouldAddEndPuncttrue
\mciteSetBstMidEndSepPunct{\mcitedefaultmidpunct}
{\mcitedefaultendpunct}{\mcitedefaultseppunct}\relax
\EndOfBibitem
\bibitem[Sivalingam \latin{et~al.}(2016)Sivalingam, Krupicka, Auer, and
  Neese]{Sivalingam:2016hr}
Sivalingam,~K.; Krupicka,~M.; Auer,~A.~A.; Neese,~F. {Comparison of fully
  internally and strongly contracted multireference configuration interaction
  procedures}. \emph{J. Chem. Phys.} \textbf{2016}, \emph{145}, 054104\relax
\mciteBstWouldAddEndPuncttrue
\mciteSetBstMidEndSepPunct{\mcitedefaultmidpunct}
{\mcitedefaultendpunct}{\mcitedefaultseppunct}\relax
\EndOfBibitem
\bibitem[{\v C}{\'\i}{\v z}ek(1969)]{Cizek:1969hv}
{\v C}{\'\i}{\v z}ek,~J. {On the Use of the Cluster Expansion and the Technique
  of Diagrams in Calculations of Correlation Effects in Atoms and Molecules}.
  \emph{Adv. Chem. Phys.} \textbf{1969}, \emph{14}, 35--89\relax
\mciteBstWouldAddEndPuncttrue
\mciteSetBstMidEndSepPunct{\mcitedefaultmidpunct}
{\mcitedefaultendpunct}{\mcitedefaultseppunct}\relax
\EndOfBibitem
\bibitem[Jeziorski and Monkhorst(1981)Jeziorski, and
  Monkhorst]{Jeziorski:1981gz}
Jeziorski,~B.; Monkhorst,~H.~J. {Coupled-cluster method for multideterminantal
  reference states}. \emph{Phys. Rev. A} \textbf{1981}, \emph{24},
  1668--1681\relax
\mciteBstWouldAddEndPuncttrue
\mciteSetBstMidEndSepPunct{\mcitedefaultmidpunct}
{\mcitedefaultendpunct}{\mcitedefaultseppunct}\relax
\EndOfBibitem
\bibitem[Haque and Mukherjee(1984)Haque, and Mukherjee]{Haque:1984dk}
Haque,~M.~A.; Mukherjee,~D. {Application of cluster expansion techniques to
  open shells: Calculation of difference energies}. \emph{J. Chem. Phys.}
  \textbf{1984}, \emph{80}, 5058--5069\relax
\mciteBstWouldAddEndPuncttrue
\mciteSetBstMidEndSepPunct{\mcitedefaultmidpunct}
{\mcitedefaultendpunct}{\mcitedefaultseppunct}\relax
\EndOfBibitem
\bibitem[Stolarczyk and Monkhorst(1985)Stolarczyk, and
  Monkhorst]{Stolarczyk:1985fk}
Stolarczyk,~L.~Z.; Monkhorst,~H.~J. {Coupled-cluster method in Fock space. I.
  General formalism}. \emph{Phys. Rev. A} \textbf{1985}, \emph{32},
  725--742\relax
\mciteBstWouldAddEndPuncttrue
\mciteSetBstMidEndSepPunct{\mcitedefaultmidpunct}
{\mcitedefaultendpunct}{\mcitedefaultseppunct}\relax
\EndOfBibitem
\bibitem[Malrieu \latin{et~al.}(1985)Malrieu, Durand, and
  Daudey]{Malrieu:1985ce}
Malrieu,~J.~P.; Durand,~P.; Daudey,~J.~P. {Intermediate Hamiltonians as a new
  class of effective Hamiltonians}. \emph{J. Phys. A: Math. Gen.}
  \textbf{1985}, \emph{18}, 809--826\relax
\mciteBstWouldAddEndPuncttrue
\mciteSetBstMidEndSepPunct{\mcitedefaultmidpunct}
{\mcitedefaultendpunct}{\mcitedefaultseppunct}\relax
\EndOfBibitem
\bibitem[Lindgren and Mukherjee(1987)Lindgren, and Mukherjee]{Lindgren:1987in}
Lindgren,~I.; Mukherjee,~D. {On the connectivity criteria in the open-shell
  coupled-cluster theory for general model spaces}. \emph{Phys. Rep.}
  \textbf{1987}, \emph{151}, 93--127\relax
\mciteBstWouldAddEndPuncttrue
\mciteSetBstMidEndSepPunct{\mcitedefaultmidpunct}
{\mcitedefaultendpunct}{\mcitedefaultseppunct}\relax
\EndOfBibitem
\bibitem[Piecuch and Paldus(1992)Piecuch, and Paldus]{Piecuch:1992gv}
Piecuch,~P.; Paldus,~J. {Orthogonally spin-adapted multi-reference Hilbert
  space coupled-cluster formalism: diagrammatic formulation}. \emph{Theor.
  Chim. Acta} \textbf{1992}, \emph{83}, 69--103\relax
\mciteBstWouldAddEndPuncttrue
\mciteSetBstMidEndSepPunct{\mcitedefaultmidpunct}
{\mcitedefaultendpunct}{\mcitedefaultseppunct}\relax
\EndOfBibitem
\bibitem[Paldus \latin{et~al.}(1993)Paldus, Piecuch, Pylypow, and
  Jeziorski]{Paldus:1993dx}
Paldus,~J.; Piecuch,~P.; Pylypow,~L.; Jeziorski,~B. {Application of
  Hilbert-space coupled-cluster theory to simple (H$_{2}$)$_{2}$ model systems:
  Planar models}. \emph{Phys. Rev. A} \textbf{1993}, \emph{47},
  2738--2782\relax
\mciteBstWouldAddEndPuncttrue
\mciteSetBstMidEndSepPunct{\mcitedefaultmidpunct}
{\mcitedefaultendpunct}{\mcitedefaultseppunct}\relax
\EndOfBibitem
\bibitem[M{\'a}{\v s}ik and Huba{\v c}(1998)M{\'a}{\v s}ik, and Huba{\v
  c}]{Masik:1998gk}
M{\'a}{\v s}ik,~J.; Huba{\v c},~I. {Multireference Brillouin-Wigner
  Coupled-Cluster Theory. Single-root approach.} \emph{Adv. Quantum Chem.}
  \textbf{1998}, \emph{31}, 75--104\relax
\mciteBstWouldAddEndPuncttrue
\mciteSetBstMidEndSepPunct{\mcitedefaultmidpunct}
{\mcitedefaultendpunct}{\mcitedefaultseppunct}\relax
\EndOfBibitem
\bibitem[Mahapatra \latin{et~al.}(1998)Mahapatra, Datta, Bandyopadhyay, and
  Mukherjee]{Mahapatra:1998cp}
Mahapatra,~U.~S.; Datta,~B.; Bandyopadhyay,~B.; Mukherjee,~D. {State-Specific
  Multi-Reference Coupled Cluster Formulations: Two Paradigms}. \emph{Adv.
  Quantum Chem.} \textbf{1998}, \emph{30}, 163--193\relax
\mciteBstWouldAddEndPuncttrue
\mciteSetBstMidEndSepPunct{\mcitedefaultmidpunct}
{\mcitedefaultendpunct}{\mcitedefaultseppunct}\relax
\EndOfBibitem
\bibitem[Mahapatra \latin{et~al.}(1998)Mahapatra, Datta, and
  Mukherjee]{Mahapatra:1998kj}
Mahapatra,~U.~S.; Datta,~B.; Mukherjee,~D. {A state-specific multi-reference
  coupled cluster formalism with molecular applications}. \emph{Mol. Phys.}
  \textbf{1998}, \emph{94}, 157--171\relax
\mciteBstWouldAddEndPuncttrue
\mciteSetBstMidEndSepPunct{\mcitedefaultmidpunct}
{\mcitedefaultendpunct}{\mcitedefaultseppunct}\relax
\EndOfBibitem
\bibitem[Mahapatra \latin{et~al.}(1999)Mahapatra, Datta, and
  Mukherjee]{Mahapatra:1999ev}
Mahapatra,~U.~S.; Datta,~B.; Mukherjee,~D. {A size-consistent state-specific
  multireference coupled cluster theory: Formal developments and molecular
  applications}. \emph{J. Chem. Phys.} \textbf{1999}, \emph{110},
  6171--6188\relax
\mciteBstWouldAddEndPuncttrue
\mciteSetBstMidEndSepPunct{\mcitedefaultmidpunct}
{\mcitedefaultendpunct}{\mcitedefaultseppunct}\relax
\EndOfBibitem
\bibitem[Li(2004)]{Li:2004ko}
Li,~S. {Block-correlated coupled cluster theory: The general formulation and
  its application to the antiferromagnetic Heisenberg model}. \emph{J. Chem.
  Phys.} \textbf{2004}, \emph{120}, 5017--5026\relax
\mciteBstWouldAddEndPuncttrue
\mciteSetBstMidEndSepPunct{\mcitedefaultmidpunct}
{\mcitedefaultendpunct}{\mcitedefaultseppunct}\relax
\EndOfBibitem
\bibitem[Hanrath(2005)]{Hanrath:2005kj}
Hanrath,~M. {An exponential multireference wave-function Ansatz}. \emph{J.
  Chem. Phys.} \textbf{2005}, \emph{123}, 084102\relax
\mciteBstWouldAddEndPuncttrue
\mciteSetBstMidEndSepPunct{\mcitedefaultmidpunct}
{\mcitedefaultendpunct}{\mcitedefaultseppunct}\relax
\EndOfBibitem
\bibitem[Evangelista and Gauss(2011)Evangelista, and Gauss]{Evangelista:2011eh}
Evangelista,~F.~A.; Gauss,~J. {An orbital-invariant internally contracted
  multireference coupled cluster approach.} \emph{J. Chem. Phys.}
  \textbf{2011}, \emph{134}, 114102\relax
\mciteBstWouldAddEndPuncttrue
\mciteSetBstMidEndSepPunct{\mcitedefaultmidpunct}
{\mcitedefaultendpunct}{\mcitedefaultseppunct}\relax
\EndOfBibitem
\bibitem[Hanauer and K{\"o}hn(2011)Hanauer, and K{\"o}hn]{Hanauer:2011ey}
Hanauer,~M.; K{\"o}hn,~A. {Pilot applications of internally contracted
  multireference coupled cluster theory, and how to choose the cluster operator
  properly}. \emph{J. Chem. Phys.} \textbf{2011}, \emph{134}, 204111\relax
\mciteBstWouldAddEndPuncttrue
\mciteSetBstMidEndSepPunct{\mcitedefaultmidpunct}
{\mcitedefaultendpunct}{\mcitedefaultseppunct}\relax
\EndOfBibitem
\bibitem[Datta \latin{et~al.}(2011)Datta, Kong, and Nooijen]{Datta:2011ca}
Datta,~D.; Kong,~L.; Nooijen,~M. {A state-specific partially internally
  contracted multireference coupled cluster approach}. \emph{J. Chem. Phys.}
  \textbf{2011}, \emph{134}, 214116\relax
\mciteBstWouldAddEndPuncttrue
\mciteSetBstMidEndSepPunct{\mcitedefaultmidpunct}
{\mcitedefaultendpunct}{\mcitedefaultseppunct}\relax
\EndOfBibitem
\bibitem[Chen and Hoffmann(2012)Chen, and Hoffmann]{Chen:2012bm}
Chen,~Z.; Hoffmann,~M.~R. {Orbitally invariant internally contracted
  multireference unitary coupled cluster theory and its perturbative
  approximation: Theory and test calculations of second order approximation}.
  \emph{J. Chem. Phys.} \textbf{2012}, \emph{137}, 014108\relax
\mciteBstWouldAddEndPuncttrue
\mciteSetBstMidEndSepPunct{\mcitedefaultmidpunct}
{\mcitedefaultendpunct}{\mcitedefaultseppunct}\relax
\EndOfBibitem
\bibitem[Hoffmann and Simons(1988)Hoffmann, and Simons]{Hoffmann:1988fs}
Hoffmann,~M.~R.; Simons,~J. {A unitary multiconfigurational coupled-cluster
  method: Theory and applications}. \emph{J. Chem. Phys.} \textbf{1988},
  \emph{88}, 993--1002\relax
\mciteBstWouldAddEndPuncttrue
\mciteSetBstMidEndSepPunct{\mcitedefaultmidpunct}
{\mcitedefaultendpunct}{\mcitedefaultseppunct}\relax
\EndOfBibitem
\bibitem[Bartlett \latin{et~al.}(1989)Bartlett, Kucharski, and
  Noga]{Bartlett:1989iz}
Bartlett,~R.~J.; Kucharski,~S.~A.; Noga,~J. {Alternative coupled-cluster
  ans{\"a}tze II. The unitary coupled-cluster method}. \emph{Chem. Phys. Lett.}
  \textbf{1989}, \emph{155}, 133--140\relax
\mciteBstWouldAddEndPuncttrue
\mciteSetBstMidEndSepPunct{\mcitedefaultmidpunct}
{\mcitedefaultendpunct}{\mcitedefaultseppunct}\relax
\EndOfBibitem
\bibitem[Watts \latin{et~al.}(1989)Watts, Trucks, and Bartlett]{Watts:1989ht}
Watts,~J.~D.; Trucks,~G.~W.; Bartlett,~R.~J. {The unitary coupled-cluster
  approach and molecular properties. Applications of the UCC(4) method}.
  \emph{Chem. Phys. Lett.} \textbf{1989}, \emph{157}, 359--366\relax
\mciteBstWouldAddEndPuncttrue
\mciteSetBstMidEndSepPunct{\mcitedefaultmidpunct}
{\mcitedefaultendpunct}{\mcitedefaultseppunct}\relax
\EndOfBibitem
\bibitem[Kutzelnigg(1991)]{Kutzelnigg:1991iw}
Kutzelnigg,~W. {Error analysis and improvements of coupled-cluster theory}.
  \emph{Theor. Chim. Acta} \textbf{1991}, \emph{80}, 349--386\relax
\mciteBstWouldAddEndPuncttrue
\mciteSetBstMidEndSepPunct{\mcitedefaultmidpunct}
{\mcitedefaultendpunct}{\mcitedefaultseppunct}\relax
\EndOfBibitem
\bibitem[Mertins and Schirmer(1996)Mertins, and Schirmer]{Mertins:1996eu}
Mertins,~F.; Schirmer,~J. {Algebraic propagator approaches and
  intermediate-state representations. I. The biorthogonal and unitary
  coupled-cluster methods}. \emph{Phys. Rev. A} \textbf{1996}, \emph{53},
  2140--2152\relax
\mciteBstWouldAddEndPuncttrue
\mciteSetBstMidEndSepPunct{\mcitedefaultmidpunct}
{\mcitedefaultendpunct}{\mcitedefaultseppunct}\relax
\EndOfBibitem
\bibitem[Taube and Bartlett(2006)Taube, and Bartlett]{Taube:2006bi}
Taube,~A.~G.; Bartlett,~R.~J. New perspectives on unitary coupled-cluster
  theory. \emph{Int. J. Quantum Chem.} \textbf{2006}, \emph{106},
  3393--3401\relax
\mciteBstWouldAddEndPuncttrue
\mciteSetBstMidEndSepPunct{\mcitedefaultmidpunct}
{\mcitedefaultendpunct}{\mcitedefaultseppunct}\relax
\EndOfBibitem
\bibitem[Harsha \latin{et~al.}(2018)Harsha, Shiozaki, and
  Scuseria]{Harsha:2018dv}
Harsha,~G.; Shiozaki,~T.; Scuseria,~G.~E. {On the difference between
  variational and unitary coupled cluster theories}. \emph{J. Chem. Phys.}
  \textbf{2018}, \emph{148}, 044107\relax
\mciteBstWouldAddEndPuncttrue
\mciteSetBstMidEndSepPunct{\mcitedefaultmidpunct}
{\mcitedefaultendpunct}{\mcitedefaultseppunct}\relax
\EndOfBibitem
\bibitem[Yung \latin{et~al.}(2014)Yung, Casanova, Mezzacapo, McClean, Lamata,
  Aspuru-Guzik, and Solano]{Yung:2014iv}
Yung,~M.~H.; Casanova,~J.; Mezzacapo,~A.; McClean,~J.; Lamata,~L.;
  Aspuru-Guzik,~A.; Solano,~E. {From transistor to trapped-ion computers for
  quantum chemistry}. \emph{Sci Rep} \textbf{2014}, \emph{4}, 714\relax
\mciteBstWouldAddEndPuncttrue
\mciteSetBstMidEndSepPunct{\mcitedefaultmidpunct}
{\mcitedefaultendpunct}{\mcitedefaultseppunct}\relax
\EndOfBibitem
\bibitem[Peruzzo \latin{et~al.}(2014)Peruzzo, McClean, Shadbolt, Yung, Zhou,
  Love, Aspuru-Guzik, and O{\textquoteright}Brien]{Peruzzo:2014kc}
Peruzzo,~A.; McClean,~J.; Shadbolt,~P.; Yung,~M.-H.; Zhou,~X.-Q.; Love,~P.~J.;
  Aspuru-Guzik,~A.; O{\textquoteright}Brien,~J.~L. {A variational eigenvalue
  solver on a photonic quantum processor}. \emph{Nat Commun} \textbf{2014},
  \emph{5}, 36\relax
\mciteBstWouldAddEndPuncttrue
\mciteSetBstMidEndSepPunct{\mcitedefaultmidpunct}
{\mcitedefaultendpunct}{\mcitedefaultseppunct}\relax
\EndOfBibitem
\bibitem[O{\textquoteright}Malley \latin{et~al.}(2016)O{\textquoteright}Malley,
  Babbush, Kivlichan, Romero, McClean, Barends, Kelly, Roushan, Tranter, Ding,
  Campbell, Chen, Chen, Chiaro, Dunsworth, Fowler, Jeffrey, Lucero, Megrant,
  Mutus, Neeley, Neill, Quintana, Sank, Vainsencher, Wenner, White, Coveney,
  Love, Neven, Aspuru-Guzik, and Martinis]{OMalley:2016dc}
O{\textquoteright}Malley,~P. J.~J.; Babbush,~R.; Kivlichan,~I.~D.; Romero,~J.;
  McClean,~J.~R.; Barends,~R.; Kelly,~J.; Roushan,~P.; Tranter,~A.; Ding,~N.;
  Campbell,~B.; Chen,~Y.; Chen,~Z.; Chiaro,~B.; Dunsworth,~A.; Fowler,~A.~G.;
  Jeffrey,~E.; Lucero,~E.; Megrant,~A.; Mutus,~J.~Y.; Neeley,~M.; Neill,~C.;
  Quintana,~C.; Sank,~D.; Vainsencher,~A.; Wenner,~J.; White,~T.~C.;
  Coveney,~P.~V.; Love,~P.~J.; Neven,~H.; Aspuru-Guzik,~A.; Martinis,~J.~M.
  {Scalable Quantum Simulation of Molecular Energies}. \emph{Phys. Rev. X}
  \textbf{2016}, \emph{6}, 361\relax
\mciteBstWouldAddEndPuncttrue
\mciteSetBstMidEndSepPunct{\mcitedefaultmidpunct}
{\mcitedefaultendpunct}{\mcitedefaultseppunct}\relax
\EndOfBibitem
\bibitem[Shen \latin{et~al.}(2017)Shen, Zhang, Zhang, Zhang, Yung, and
  Kim]{Shen:2017cc}
Shen,~Y.; Zhang,~X.; Zhang,~S.; Zhang,~J.-N.; Yung,~M.-H.; Kim,~K. {Quantum
  implementation of the unitary coupled cluster for simulating molecular
  electronic structure}. \emph{Phys. Rev. A} \textbf{2017}, \emph{95},
  020501\relax
\mciteBstWouldAddEndPuncttrue
\mciteSetBstMidEndSepPunct{\mcitedefaultmidpunct}
{\mcitedefaultendpunct}{\mcitedefaultseppunct}\relax
\EndOfBibitem
\bibitem[Dumitrescu \latin{et~al.}(2018)Dumitrescu, McCaskey, Hagen, Jansen,
  Morris, Papenbrock, Pooser, Dean, and Lougovski]{Dumitrescu:2018fu}
Dumitrescu,~E.~F.; McCaskey,~A.~J.; Hagen,~G.; Jansen,~G.~R.; Morris,~T.~D.;
  Papenbrock,~T.; Pooser,~R.~C.; Dean,~D.~J.; Lougovski,~P. {Cloud Quantum
  Computing of an Atomic Nucleus}. \emph{Phys. Rev. Lett.} \textbf{2018},
  \emph{120}, 210501\relax
\mciteBstWouldAddEndPuncttrue
\mciteSetBstMidEndSepPunct{\mcitedefaultmidpunct}
{\mcitedefaultendpunct}{\mcitedefaultseppunct}\relax
\EndOfBibitem
\bibitem[Ryabinkin \latin{et~al.}(2018)Ryabinkin, Yen, Genin, and
  Izmaylov]{Ryabinkin:2018jw}
Ryabinkin,~I.~G.; Yen,~T.-C.; Genin,~S.~N.; Izmaylov,~A.~F. {Qubit Coupled
  Cluster Method: A Systematic Approach to Quantum Chemistry on a Quantum
  Computer}. \emph{J. Chem. Theory Comput.} \textbf{2018}, \emph{14},
  6317--6326\relax
\mciteBstWouldAddEndPuncttrue
\mciteSetBstMidEndSepPunct{\mcitedefaultmidpunct}
{\mcitedefaultendpunct}{\mcitedefaultseppunct}\relax
\EndOfBibitem
\bibitem[Hempel \latin{et~al.}(2018)Hempel, Maier, Romero, McClean, Monz, Shen,
  Jurcevic, Lanyon, Love, Babbush, Aspuru-Guzik, Blatt, and
  Roos]{Hempel:2018ip}
Hempel,~C.; Maier,~C.; Romero,~J.; McClean,~J.; Monz,~T.; Shen,~H.;
  Jurcevic,~P.; Lanyon,~B.~P.; Love,~P.; Babbush,~R.; Aspuru-Guzik,~A.;
  Blatt,~R.; Roos,~C.~F. {Quantum Chemistry Calculations on a Trapped-Ion
  Quantum Simulator}. \emph{Phys. Rev. X} \textbf{2018}, \emph{8}, 031022\relax
\mciteBstWouldAddEndPuncttrue
\mciteSetBstMidEndSepPunct{\mcitedefaultmidpunct}
{\mcitedefaultendpunct}{\mcitedefaultseppunct}\relax
\EndOfBibitem
\bibitem[Romero \latin{et~al.}(2019)Romero, Babbush, McClean, Hempel, Love, and
  Aspuru-Guzik]{Romero:2019hk}
Romero,~J.; Babbush,~R.; McClean,~J.~R.; Hempel,~C.; Love,~P.~J.;
  Aspuru-Guzik,~A. {Strategies for quantum computing molecular energies using
  the unitary coupled cluster ansatz}. \emph{Quantum Sci. Technol.}
  \textbf{2019}, \emph{4}, 014008\relax
\mciteBstWouldAddEndPuncttrue
\mciteSetBstMidEndSepPunct{\mcitedefaultmidpunct}
{\mcitedefaultendpunct}{\mcitedefaultseppunct}\relax
\EndOfBibitem
\bibitem[Lee \latin{et~al.}(2018)Lee, Huggins, Head-Gordon, and
  Whaley]{Lee:2018cy}
Lee,~J.; Huggins,~W.~J.; Head-Gordon,~M.; Whaley,~K.~B. {Generalized Unitary
  Coupled Cluster Wave functions for Quantum Computation}. \emph{J. Chem.
  Theory Comput.} \textbf{2018}, \emph{15}, 311--324\relax
\mciteBstWouldAddEndPuncttrue
\mciteSetBstMidEndSepPunct{\mcitedefaultmidpunct}
{\mcitedefaultendpunct}{\mcitedefaultseppunct}\relax
\EndOfBibitem
\bibitem[Li \latin{et~al.}(2019)Li, Hu, Zhang, Song, and Yung]{Li:2019id}
Li,~Y.; Hu,~J.; Zhang,~X.~M.; Song,~Z.; Yung,~M.-H. {Variational Quantum
  Simulation for Quantum Chemistry}. \emph{Adv. Theory Simul.} \textbf{2019},
  \emph{309}, 1800182\relax
\mciteBstWouldAddEndPuncttrue
\mciteSetBstMidEndSepPunct{\mcitedefaultmidpunct}
{\mcitedefaultendpunct}{\mcitedefaultseppunct}\relax
\EndOfBibitem
\bibitem[K{\" u}hn \latin{et~al.}(2018)K{\" u}hn, Zanker, Deglmann, Marthaler,
  and Wei{\ss}]{Kuhn:2018Accuracy}
K{\" u}hn,~M.; Zanker,~S.; Deglmann,~P.; Marthaler,~M.; Wei{\ss},~H. {Accuracy
  and Resource Estimations for Quantum Chemistry on a Near-term Quantum
  Computer}. \emph{\tt arXiv:1812.06814 [quant-ph]} \textbf{2018}, \relax
\mciteBstWouldAddEndPunctfalse
\mciteSetBstMidEndSepPunct{\mcitedefaultmidpunct}
{}{\mcitedefaultseppunct}\relax
\EndOfBibitem
\bibitem[Nam \latin{et~al.}(2019)Nam, Chen, Pisenti, Wright, Delaney, Maslov,
  Brown, Allen, Amini, Apisdorf, Beck, Blinov, Chaplin, Chmielewski, Collins,
  Debnath, Ducore, Hudek, Keesan, Kreikemeier, Mizrahi, Solomon, Williams,
  Wong-Campos, Monroe, and Kim]{Nam:2019Ground}
Nam,~Y.; Chen,~J.-S.; Pisenti,~N.~C.; Wright,~K.; Delaney,~C.; Maslov,~D.;
  Brown,~K.~R.; Allen,~S.; Amini,~J.~M.; Apisdorf,~J.; Beck,~K.~M.; Blinov,~A.;
  Chaplin,~V.; Chmielewski,~M.; Collins,~C.; Debnath,~S.; Ducore,~A.~M.;
  Hudek,~K.~M.; Keesan,~M.; Kreikemeier,~S.~M.; Mizrahi,~J.; Solomon,~P.;
  Williams,~M.; Wong-Campos,~J.~D.; Monroe,~C.; Kim,~J. {Ground-state energy
  estimation of the water molecule on a trapped ion quantum computer}.
  \emph{\tt arXiv:1902.10171v2 [quant-ph]} \textbf{2019}, \relax
\mciteBstWouldAddEndPunctfalse
\mciteSetBstMidEndSepPunct{\mcitedefaultmidpunct}
{}{\mcitedefaultseppunct}\relax
\EndOfBibitem
\bibitem[Bauman \latin{et~al.}(2019)Bauman, Bylaska, Krishnamoorthy, Low,
  Wiebe, and Kowalski]{Bauman:2019wj}
Bauman,~N.~P.; Bylaska,~E.~J.; Krishnamoorthy,~S.; Low,~G.~H.; Wiebe,~N.;
  Kowalski,~K. {Downfolding of many-body Hamiltonians using active-space
  models: extension of the sub-system embedding sub-algebras approach to
  unitary coupled cluster formalisms}. \emph{\tt arXiv:1902.01553v2 [quant-ph]}
  \textbf{2019}, \relax
\mciteBstWouldAddEndPunctfalse
\mciteSetBstMidEndSepPunct{\mcitedefaultmidpunct}
{}{\mcitedefaultseppunct}\relax
\EndOfBibitem
\bibitem[Evangelista(2011)]{Evangelista:2011jp}
Evangelista,~F.~A. {Alternative single-reference coupled cluster approaches for
  multireference problems: the simpler, the better.} \emph{J. Chem. Phys.}
  \textbf{2011}, \emph{134}, 224102\relax
\mciteBstWouldAddEndPuncttrue
\mciteSetBstMidEndSepPunct{\mcitedefaultmidpunct}
{\mcitedefaultendpunct}{\mcitedefaultseppunct}\relax
\EndOfBibitem
\bibitem[Hanauer and K{\"o}hn(2012)Hanauer, and K{\"o}hn]{Hanauer:2012gf}
Hanauer,~M.; K{\"o}hn,~A. {Perturbative treatment of triple excitations in
  internally contracted multireference coupled cluster theory.} \emph{J. Chem.
  Phys.} \textbf{2012}, \emph{136}, 204107\relax
\mciteBstWouldAddEndPuncttrue
\mciteSetBstMidEndSepPunct{\mcitedefaultmidpunct}
{\mcitedefaultendpunct}{\mcitedefaultseppunct}\relax
\EndOfBibitem
\bibitem[Yanai and Chan(2006)Yanai, and Chan]{Yanai:2006gi}
Yanai,~T.; Chan,~G. K.-L. {Canonical transformation theory for multireference
  problems}. \emph{J. Chem. Phys.} \textbf{2006}, \emph{124}, 194106\relax
\mciteBstWouldAddEndPuncttrue
\mciteSetBstMidEndSepPunct{\mcitedefaultmidpunct}
{\mcitedefaultendpunct}{\mcitedefaultseppunct}\relax
\EndOfBibitem
\bibitem[Yanai and Chan(2007)Yanai, and Chan]{Yanai:2007ix}
Yanai,~T.; Chan,~G. K.-L. {Canonical transformation theory from extended normal
  ordering}. \emph{J. Chem. Phys.} \textbf{2007}, \emph{127}, 104107\relax
\mciteBstWouldAddEndPuncttrue
\mciteSetBstMidEndSepPunct{\mcitedefaultmidpunct}
{\mcitedefaultendpunct}{\mcitedefaultseppunct}\relax
\EndOfBibitem
\bibitem[Wegner(1994)]{Wegner:1994kh}
Wegner,~F. {Flow-equations for Hamiltonians}. \emph{Ann. Phys.} \textbf{1994},
  \emph{506}, 77--91\relax
\mciteBstWouldAddEndPuncttrue
\mciteSetBstMidEndSepPunct{\mcitedefaultmidpunct}
{\mcitedefaultendpunct}{\mcitedefaultseppunct}\relax
\EndOfBibitem
\bibitem[Kehrein(2006)]{Kehrein:2006vz}
Kehrein,~S. \emph{The Flow Equation Approach to Many-Particle Systems};
  Springer Berlin Heidelberg, 2006\relax
\mciteBstWouldAddEndPuncttrue
\mciteSetBstMidEndSepPunct{\mcitedefaultmidpunct}
{\mcitedefaultendpunct}{\mcitedefaultseppunct}\relax
\EndOfBibitem
\bibitem[Evangelista(2014)]{Evangelista:2014kt}
Evangelista,~F.~A. {A driven similarity renormalization group approach to
  quantum many-body problems.} \emph{J. Chem. Phys.} \textbf{2014}, \emph{141},
  054109\relax
\mciteBstWouldAddEndPuncttrue
\mciteSetBstMidEndSepPunct{\mcitedefaultmidpunct}
{\mcitedefaultendpunct}{\mcitedefaultseppunct}\relax
\EndOfBibitem
\bibitem[Li and Evangelista(2015)Li, and Evangelista]{Li:2015iz}
Li,~C.; Evangelista,~F.~A. Multireference Driven Similarity Renormalization
  Group: A Second-Order Perturbative Analysis. \emph{J. Chem. Theory Comput.}
  \textbf{2015}, \emph{11}, 2097--2108\relax
\mciteBstWouldAddEndPuncttrue
\mciteSetBstMidEndSepPunct{\mcitedefaultmidpunct}
{\mcitedefaultendpunct}{\mcitedefaultseppunct}\relax
\EndOfBibitem
\bibitem[Li and Evangelista(2016)Li, and Evangelista]{Li:2016hb}
Li,~C.; Evangelista,~F.~A. Towards numerically robust multireference theories:
  The driven similarity renormalization group truncated to one- and two-body
  operators. \emph{J. Chem. Phys.} \textbf{2016}, \emph{144}, 164114\relax
\mciteBstWouldAddEndPuncttrue
\mciteSetBstMidEndSepPunct{\mcitedefaultmidpunct}
{\mcitedefaultendpunct}{\mcitedefaultseppunct}\relax
\EndOfBibitem
\bibitem[Li and Evangelista(2018)Li, and Evangelista]{Li:2018dy}
Li,~C.; Evangelista,~F.~A. {Erratum: {\textquotedblleft}Towards numerically
  robust multireference theories: The driven similarity renormalization group
  truncated to one- and two-body operators{\textquotedblright} [J. Chem. Phys.
  144, 164114 (2016)]}. \emph{J. Chem. Phys.} \textbf{2018}, \emph{148},
  079903\relax
\mciteBstWouldAddEndPuncttrue
\mciteSetBstMidEndSepPunct{\mcitedefaultmidpunct}
{\mcitedefaultendpunct}{\mcitedefaultseppunct}\relax
\EndOfBibitem
\bibitem[Li and Evangelista(2018)Li, and Evangelista]{Li:2018kl}
Li,~C.; Evangelista,~F.~A. {Driven similarity renormalization group for excited
  states: A state-averaged perturbation theory}. \emph{J. Chem. Phys.}
  \textbf{2018}, \emph{148}, 124106\relax
\mciteBstWouldAddEndPuncttrue
\mciteSetBstMidEndSepPunct{\mcitedefaultmidpunct}
{\mcitedefaultendpunct}{\mcitedefaultseppunct}\relax
\EndOfBibitem
\bibitem[Li and Evangelista(2019)Li, and Evangelista]{Li:2019xxx}
Li,~C.; Evangelista,~F.~A. Multireference Theories of Electron Correlation
  Based on the Driven Similarity Renormalization Group. \emph{Annu. Rev. Phys.
  Chem.} \textbf{2019}, \emph{70}, in press\relax
\mciteBstWouldAddEndPuncttrue
\mciteSetBstMidEndSepPunct{\mcitedefaultmidpunct}
{\mcitedefaultendpunct}{\mcitedefaultseppunct}\relax
\EndOfBibitem
\bibitem[Meller \latin{et~al.}(1996)Meller, Malrieu, and
  Caballol]{Meller:1996ey}
Meller,~J.; Malrieu,~J.~P.; Caballol,~R. {State-specific coupled cluster-type
  dressing of multireference singles and doubles configuration interaction
  matrix}. \emph{J. Chem. Phys.} \textbf{1996}, \emph{104}, 4068--4076\relax
\mciteBstWouldAddEndPuncttrue
\mciteSetBstMidEndSepPunct{\mcitedefaultmidpunct}
{\mcitedefaultendpunct}{\mcitedefaultseppunct}\relax
\EndOfBibitem
\bibitem[K{\"o}hn \latin{et~al.}(2013)K{\"o}hn, Hanauer, M{\"u}ck, Jagau, and
  Gauss]{Koehn:2013cp}
K{\"o}hn,~A.; Hanauer,~M.; M{\"u}ck,~L.~A.; Jagau,~T.-C.; Gauss,~J.
  {State-specific multireference coupled-cluster theory}. \emph{Wiley
  Interdiscip. Rev.: Comput. Mol. Sci.} \textbf{2013}, \emph{3}, 176--197\relax
\mciteBstWouldAddEndPuncttrue
\mciteSetBstMidEndSepPunct{\mcitedefaultmidpunct}
{\mcitedefaultendpunct}{\mcitedefaultseppunct}\relax
\EndOfBibitem
\bibitem[Lyakh \latin{et~al.}(2012)Lyakh, Musia{\l}, Lotrich, and
  Bartlett]{Lyakh:2012cn}
Lyakh,~D.~I.; Musia{\l},~M.; Lotrich,~V.~F.; Bartlett,~R.~J. {Multireference
  nature of chemistry: the coupled-cluster view.} \emph{Chem. Rev.}
  \textbf{2012}, \emph{112}, 182--243\relax
\mciteBstWouldAddEndPuncttrue
\mciteSetBstMidEndSepPunct{\mcitedefaultmidpunct}
{\mcitedefaultendpunct}{\mcitedefaultseppunct}\relax
\EndOfBibitem
\bibitem[Evangelista(2018)]{Evangelista:2018bt}
Evangelista,~F.~A. {Perspective: Multireference coupled cluster theories of
  dynamical electron correlation}. \emph{J. Chem. Phys.} \textbf{2018},
  \emph{149}, 030901\relax
\mciteBstWouldAddEndPuncttrue
\mciteSetBstMidEndSepPunct{\mcitedefaultmidpunct}
{\mcitedefaultendpunct}{\mcitedefaultseppunct}\relax
\EndOfBibitem
\bibitem[Schucan and Weidenm{\"u}ller(1972)Schucan, and
  Weidenm{\"u}ller]{Schucan:1972bs}
Schucan,~T.~H.; Weidenm{\"u}ller,~H.~A. {The effective interaction in nuclei
  and its perturbation expansion: An algebraic approach}. \emph{Ann. Phys.}
  \textbf{1972}, \emph{73}, 108--135\relax
\mciteBstWouldAddEndPuncttrue
\mciteSetBstMidEndSepPunct{\mcitedefaultmidpunct}
{\mcitedefaultendpunct}{\mcitedefaultseppunct}\relax
\EndOfBibitem
\bibitem[Schucan and Weidenm{\"u}ller(1973)Schucan, and
  Weidenm{\"u}ller]{Schucan:1973dy}
Schucan,~T.~H.; Weidenm{\"u}ller,~H.~A. {Perturbation theory for the effective
  interaction in nuclei}. \emph{Ann. Phys.} \textbf{1973}, \emph{76},
  483--509\relax
\mciteBstWouldAddEndPuncttrue
\mciteSetBstMidEndSepPunct{\mcitedefaultmidpunct}
{\mcitedefaultendpunct}{\mcitedefaultseppunct}\relax
\EndOfBibitem
\bibitem[Salomonson \latin{et~al.}(1980)Salomonson, Lindgren, and
  M{\aa}rtensson]{Salomonson:1980ko}
Salomonson,~S.; Lindgren,~I.; M{\aa}rtensson,~A.-M. {Numerical Many-Body
  Perturbation Calculations on Be-like Systems Using a Multi-Configurational
  Model Space}. \emph{Phys. Scr.} \textbf{1980}, \emph{21}, 351--356\relax
\mciteBstWouldAddEndPuncttrue
\mciteSetBstMidEndSepPunct{\mcitedefaultmidpunct}
{\mcitedefaultendpunct}{\mcitedefaultseppunct}\relax
\EndOfBibitem
\bibitem[Evangelisti \latin{et~al.}(1987)Evangelisti, Daudey, and
  Malrieu]{Evangelisti:1987fw}
Evangelisti,~S.; Daudey,~J.~P.; Malrieu,~J.~P. {Qualitative intruder-state
  problems in effective Hamiltonian theory and their solution through
  intermediate Hamiltonians}. \emph{Phys. Rev. A} \textbf{1987}, \emph{35},
  4930--4941\relax
\mciteBstWouldAddEndPuncttrue
\mciteSetBstMidEndSepPunct{\mcitedefaultmidpunct}
{\mcitedefaultendpunct}{\mcitedefaultseppunct}\relax
\EndOfBibitem
\bibitem[Zarrabian \latin{et~al.}(1990)Zarrabian, Laidig, and
  Bartlett]{Zarrabian:1990ig}
Zarrabian,~S.; Laidig,~W.~D.; Bartlett,~R.~J. {Convergence properties of
  multireference many-body perturbation theory}. \emph{Phys. Rev. A}
  \textbf{1990}, \emph{41}, 4711--4720\relax
\mciteBstWouldAddEndPuncttrue
\mciteSetBstMidEndSepPunct{\mcitedefaultmidpunct}
{\mcitedefaultendpunct}{\mcitedefaultseppunct}\relax
\EndOfBibitem
\bibitem[Kowalski and Piecuch(2000)Kowalski, and Piecuch]{Kowalski:2000cj}
Kowalski,~K.; Piecuch,~P. {Complete set of solutions of multireference
  coupled-cluster equations: The state-universal formalism}. \emph{Phys. Rev.
  A} \textbf{2000}, \emph{61}, 052506\relax
\mciteBstWouldAddEndPuncttrue
\mciteSetBstMidEndSepPunct{\mcitedefaultmidpunct}
{\mcitedefaultendpunct}{\mcitedefaultseppunct}\relax
\EndOfBibitem
\bibitem[Kowalski and Piecuch(2000)Kowalski, and Piecuch]{Kowalski:2000cv}
Kowalski,~K.; Piecuch,~P. {Complete set of solutions of the generalized Bloch
  equation}. \emph{Int. J. Quantum Chem.} \textbf{2000}, \emph{80},
  757--781\relax
\mciteBstWouldAddEndPuncttrue
\mciteSetBstMidEndSepPunct{\mcitedefaultmidpunct}
{\mcitedefaultendpunct}{\mcitedefaultseppunct}\relax
\EndOfBibitem
\bibitem[Li and Evangelista(2017)Li, and Evangelista]{Li:2017bx}
Li,~C.; Evangelista,~F.~A. {Driven similarity renormalization group:
  Third-order multireference perturbation theory}. \emph{J. Chem. Phys.}
  \textbf{2017}, \emph{146}, 124132\relax
\mciteBstWouldAddEndPuncttrue
\mciteSetBstMidEndSepPunct{\mcitedefaultmidpunct}
{\mcitedefaultendpunct}{\mcitedefaultseppunct}\relax
\EndOfBibitem
\bibitem[Li and Evangelista(2018)Li, and Evangelista]{Li:2018fn}
Li,~C.; Evangelista,~F.~A. {Erratum: {\textquotedblleft}Driven similarity
  renormalization group: Third-order multireference perturbation
  theory{\textquotedblright} [J. Chem. Phys. 146, 124132 (2017)]}. \emph{J.
  Chem. Phys.} \textbf{2018}, \emph{148}, 079902\relax
\mciteBstWouldAddEndPuncttrue
\mciteSetBstMidEndSepPunct{\mcitedefaultmidpunct}
{\mcitedefaultendpunct}{\mcitedefaultseppunct}\relax
\EndOfBibitem
\bibitem[Evangelista \latin{et~al.}(2012)Evangelista, Hanauer, K{\"o}hn, and
  Gauss]{Evangelista:2012fo}
Evangelista,~F.~A.; Hanauer,~M.; K{\"o}hn,~A.; Gauss,~J. {A sequential
  transformation approach to the internally contracted multireference coupled
  cluster method.} \emph{J. Chem. Phys.} \textbf{2012}, \emph{136},
  204108\relax
\mciteBstWouldAddEndPuncttrue
\mciteSetBstMidEndSepPunct{\mcitedefaultmidpunct}
{\mcitedefaultendpunct}{\mcitedefaultseppunct}\relax
\EndOfBibitem
\bibitem[Whitten(1973)]{Whitten:1973ju}
Whitten,~J.~L. {Coulombic potential energy integrals and approximations}.
  \emph{J. Chem. Phys.} \textbf{1973}, \emph{58}, 4496--4501\relax
\mciteBstWouldAddEndPuncttrue
\mciteSetBstMidEndSepPunct{\mcitedefaultmidpunct}
{\mcitedefaultendpunct}{\mcitedefaultseppunct}\relax
\EndOfBibitem
\bibitem[Dunlap \latin{et~al.}(1979)Dunlap, Connolly, and Sabin]{Dunlap:1979gh}
Dunlap,~B.~I.; Connolly,~J. W.~D.; Sabin,~J.~R. {On some approximations in
  applications of X$\alpha$ theory}. \emph{J. Chem. Phys.} \textbf{1979},
  \emph{71}, 3396--3402\relax
\mciteBstWouldAddEndPuncttrue
\mciteSetBstMidEndSepPunct{\mcitedefaultmidpunct}
{\mcitedefaultendpunct}{\mcitedefaultseppunct}\relax
\EndOfBibitem
\bibitem[Kendall and Fruchtl(1997)Kendall, and Fruchtl]{Kendall:1997kh}
Kendall,~R.~A.; Fruchtl,~H.~A. {The impact of the resolution of the identity
  approximate integral method on modern ab initio algorithm development}.
  \emph{Theor. Chem. Acc.} \textbf{1997}, \emph{97}, 158--163\relax
\mciteBstWouldAddEndPuncttrue
\mciteSetBstMidEndSepPunct{\mcitedefaultmidpunct}
{\mcitedefaultendpunct}{\mcitedefaultseppunct}\relax
\EndOfBibitem
\bibitem[Beebe and Linderberg(1977)Beebe, and Linderberg]{Beebe:1977dp}
Beebe,~N. H.~F.; Linderberg,~J. {Simplifications in the generation and
  transformation of two-electron integrals in molecular calculations}.
  \emph{Int. J. Quantum Chem.} \textbf{1977}, \emph{12}, 683--705\relax
\mciteBstWouldAddEndPuncttrue
\mciteSetBstMidEndSepPunct{\mcitedefaultmidpunct}
{\mcitedefaultendpunct}{\mcitedefaultseppunct}\relax
\EndOfBibitem
\bibitem[R{\o}eggen and Wisl{\o}ff-Nilssen(1986)R{\o}eggen, and
  Wisl{\o}ff-Nilssen]{Roeggen:1986fp}
R{\o}eggen,~I.; Wisl{\o}ff-Nilssen,~E. {On the Beebe-Linderberg two-electron
  integral approximation}. \emph{Chem. Phys. Lett.} \textbf{1986}, \emph{132},
  154--160\relax
\mciteBstWouldAddEndPuncttrue
\mciteSetBstMidEndSepPunct{\mcitedefaultmidpunct}
{\mcitedefaultendpunct}{\mcitedefaultseppunct}\relax
\EndOfBibitem
\bibitem[Koch \latin{et~al.}(2003)Koch, S{\'a}nchez~de Mer{\'a}s, and
  Pedersen]{Koch:2003go}
Koch,~H.; S{\'a}nchez~de Mer{\'a}s,~A.; Pedersen,~T.~B. {Reduced scaling in
  electronic structure calculations using Cholesky decompositions}. \emph{J.
  Chem. Phys.} \textbf{2003}, \emph{118}, 9481--9484\relax
\mciteBstWouldAddEndPuncttrue
\mciteSetBstMidEndSepPunct{\mcitedefaultmidpunct}
{\mcitedefaultendpunct}{\mcitedefaultseppunct}\relax
\EndOfBibitem
\bibitem[Boman \latin{et~al.}(2008)Boman, Koch, and de~Mer{\'a}s]{Boman:2008fw}
Boman,~L.; Koch,~H.; de~Mer{\'a}s,~A.~S. {Method specific Cholesky
  decomposition: Coulomb and exchange energies}. \emph{J. Chem. Phys.}
  \textbf{2008}, \emph{129}, 134107\relax
\mciteBstWouldAddEndPuncttrue
\mciteSetBstMidEndSepPunct{\mcitedefaultmidpunct}
{\mcitedefaultendpunct}{\mcitedefaultseppunct}\relax
\EndOfBibitem
\bibitem[Aquilante \latin{et~al.}(2009)Aquilante, Gagliardi, Pedersen, and
  Lindh]{Aquilante:2009eu}
Aquilante,~F.; Gagliardi,~L.; Pedersen,~T.~B.; Lindh,~R. {Atomic Cholesky
  decompositions: A route to unbiased auxiliary basis sets for density fitting
  approximation with tunable accuracy and efficiency}. \emph{J. Chem. Phys.}
  \textbf{2009}, \emph{130}, 154107\relax
\mciteBstWouldAddEndPuncttrue
\mciteSetBstMidEndSepPunct{\mcitedefaultmidpunct}
{\mcitedefaultendpunct}{\mcitedefaultseppunct}\relax
\EndOfBibitem
\bibitem[Parrish \latin{et~al.}(2014)Parrish, Sherrill, Hohenstein, Kokkila,
  and Mart{\'\i}nez]{Parrish:2014ig}
Parrish,~R.~M.; Sherrill,~C.~D.; Hohenstein,~E.~G.; Kokkila,~S. I.~L.;
  Mart{\'\i}nez,~T.~J. {Communication: Acceleration of coupled cluster singles
  and doubles via orbital-weighted least-squares tensor hypercontraction}.
  \emph{J. Chem. Phys.} \textbf{2014}, \emph{140}, 181102\relax
\mciteBstWouldAddEndPuncttrue
\mciteSetBstMidEndSepPunct{\mcitedefaultmidpunct}
{\mcitedefaultendpunct}{\mcitedefaultseppunct}\relax
\EndOfBibitem
\bibitem[Vahtras \latin{et~al.}(1993)Vahtras, Alml{\"o}f, and
  Feyereisen]{Vahtras:1993db}
Vahtras,~O.; Alml{\"o}f,~J.; Feyereisen,~M.~W. {Integral approximations for
  LCAO-SCF calculations}. \emph{Chem. Phys. Lett.} \textbf{1993}, \emph{213},
  514--518\relax
\mciteBstWouldAddEndPuncttrue
\mciteSetBstMidEndSepPunct{\mcitedefaultmidpunct}
{\mcitedefaultendpunct}{\mcitedefaultseppunct}\relax
\EndOfBibitem
\bibitem[Aquilante \latin{et~al.}(2007)Aquilante, Pedersen, and
  Lindh]{Aquilante:2007es}
Aquilante,~F.; Pedersen,~T.~B.; Lindh,~R. {Low-cost evaluation of the exchange
  Fock matrix from Cholesky and density fitting representations of the electron
  repulsion integrals}. \emph{J. Chem. Phys.} \textbf{2007}, \emph{126},
  194106\relax
\mciteBstWouldAddEndPuncttrue
\mciteSetBstMidEndSepPunct{\mcitedefaultmidpunct}
{\mcitedefaultendpunct}{\mcitedefaultseppunct}\relax
\EndOfBibitem
\bibitem[Feyereisen \latin{et~al.}(1993)Feyereisen, Fitzgerald, and
  Komormicki]{Feyereisen:1993ja}
Feyereisen,~M.; Fitzgerald,~G.; Komormicki,~A. {Use of approximate integrals in
  ab initio theory. An application in MP2 energy calculations}. \emph{Chem.
  Phys. Lett.} \textbf{1993}, \emph{208}, 359--363\relax
\mciteBstWouldAddEndPuncttrue
\mciteSetBstMidEndSepPunct{\mcitedefaultmidpunct}
{\mcitedefaultendpunct}{\mcitedefaultseppunct}\relax
\EndOfBibitem
\bibitem[Weigend(2002)]{Weigend:2002ic}
Weigend,~F. {A fully direct RI-HF algorithm: Implementation, optimised
  auxiliary basis sets, demonstration of accuracy and efficiency}. \emph{Phys.
  Chem. Chem. Phys.} \textbf{2002}, \emph{4}, 4285--4291\relax
\mciteBstWouldAddEndPuncttrue
\mciteSetBstMidEndSepPunct{\mcitedefaultmidpunct}
{\mcitedefaultendpunct}{\mcitedefaultseppunct}\relax
\EndOfBibitem
\bibitem[Werner \latin{et~al.}(2003)Werner, Manby, and Knowles]{Werner:2003gq}
Werner,~H.-J.; Manby,~F.~R.; Knowles,~P.~J. {Fast linear scaling second-order
  M{\o}ller-Plesset perturbation theory (MP2) using local and density fitting
  approximations}. \emph{J. Chem. Phys.} \textbf{2003}, \emph{118},
  8149--8160\relax
\mciteBstWouldAddEndPuncttrue
\mciteSetBstMidEndSepPunct{\mcitedefaultmidpunct}
{\mcitedefaultendpunct}{\mcitedefaultseppunct}\relax
\EndOfBibitem
\bibitem[Aquilante \latin{et~al.}(2008)Aquilante, Malmqvist, Pedersen, Ghosh,
  and Roos]{Aquilante:2008gk}
Aquilante,~F.; Malmqvist,~P.-{\AA}.; Pedersen,~T.~B.; Ghosh,~A.; Roos,~B.~O.
  {Cholesky Decomposition-Based Multiconfiguration Second-Order Perturbation
  Theory (CD-CASPT2): Application to the Spin-State Energetics of Co
  III(diiminato)(NPh)}. \emph{J. Chem. Theory Comput.} \textbf{2008}, \emph{4},
  694--702\relax
\mciteBstWouldAddEndPuncttrue
\mciteSetBstMidEndSepPunct{\mcitedefaultmidpunct}
{\mcitedefaultendpunct}{\mcitedefaultseppunct}\relax
\EndOfBibitem
\bibitem[Bostr{\"o}m \latin{et~al.}(2010)Bostr{\"o}m, Delcey, Aquilante,
  Serrano-Andr{\'e}s, Pedersen, and Lindh]{Bostrom:2010is}
Bostr{\"o}m,~J.; Delcey,~M.~G.; Aquilante,~F.; Serrano-Andr{\'e}s,~L.;
  Pedersen,~T.~B.; Lindh,~R. {Calibration of Cholesky Auxiliary Basis Sets for
  Multiconfigurational Perturbation Theory Calculations of Excitation
  Energies}. \emph{J. Chem. Theory Comput.} \textbf{2010}, \emph{6},
  747--754\relax
\mciteBstWouldAddEndPuncttrue
\mciteSetBstMidEndSepPunct{\mcitedefaultmidpunct}
{\mcitedefaultendpunct}{\mcitedefaultseppunct}\relax
\EndOfBibitem
\bibitem[Gy{\H o}rffy \latin{et~al.}(2013)Gy{\H o}rffy, Shiozaki, Knizia, and
  Werner]{Gyorffy:2013kf}
Gy{\H o}rffy,~W.; Shiozaki,~T.; Knizia,~G.; Werner,~H.-J. {Analytical energy
  gradients for second-order multireference perturbation theory using density
  fitting.} \emph{J. Chem. Phys.} \textbf{2013}, \emph{138}, 104104\relax
\mciteBstWouldAddEndPuncttrue
\mciteSetBstMidEndSepPunct{\mcitedefaultmidpunct}
{\mcitedefaultendpunct}{\mcitedefaultseppunct}\relax
\EndOfBibitem
\bibitem[Hannon \latin{et~al.}(2016)Hannon, Li, and Evangelista]{Hannon:2016bh}
Hannon,~K.~P.; Li,~C.; Evangelista,~F.~A. An integral-factorized implementation
  of the driven similarity renormalization group second-order multireference
  perturbation theory. \emph{J. Chem. Phys.} \textbf{2016}, \emph{144},
  204111\relax
\mciteBstWouldAddEndPuncttrue
\mciteSetBstMidEndSepPunct{\mcitedefaultmidpunct}
{\mcitedefaultendpunct}{\mcitedefaultseppunct}\relax
\EndOfBibitem
\bibitem[Freitag \latin{et~al.}(2017)Freitag, Knecht, Angeli, and
  Reiher]{Freitag:2017ir}
Freitag,~L.; Knecht,~S.; Angeli,~C.; Reiher,~M. {Multireference Perturbation
  Theory with Cholesky Decomposition for the Density Matrix Renormalization
  Group}. \emph{J. Chem. Theory Comput.} \textbf{2017}, \emph{13},
  451--459\relax
\mciteBstWouldAddEndPuncttrue
\mciteSetBstMidEndSepPunct{\mcitedefaultmidpunct}
{\mcitedefaultendpunct}{\mcitedefaultseppunct}\relax
\EndOfBibitem
\bibitem[H{\"a}ttig and Weigend(2000)H{\"a}ttig, and Weigend]{Hattig:2000cr}
H{\"a}ttig,~C.; Weigend,~F. {CC2 excitation energy calculations on large
  molecules using the resolution of the identity approximation}. \emph{J. Chem.
  Phys.} \textbf{2000}, \emph{113}, 5154--5161\relax
\mciteBstWouldAddEndPuncttrue
\mciteSetBstMidEndSepPunct{\mcitedefaultmidpunct}
{\mcitedefaultendpunct}{\mcitedefaultseppunct}\relax
\EndOfBibitem
\bibitem[Pedersen \latin{et~al.}(2004)Pedersen, de~Mer{\'a}s, and
  Koch]{Pedersen:2004kn}
Pedersen,~T.~B.; de~Mer{\'a}s,~A. M. J.~S.; Koch,~H. {Polarizability and
  optical rotation calculated from the approximate coupled cluster singles and
  doubles CC2 linear response theory using Cholesky decompositions}. \emph{J.
  Chem. Phys.} \textbf{2004}, \emph{120}, 8887--8897\relax
\mciteBstWouldAddEndPuncttrue
\mciteSetBstMidEndSepPunct{\mcitedefaultmidpunct}
{\mcitedefaultendpunct}{\mcitedefaultseppunct}\relax
\EndOfBibitem
\bibitem[Rendell and Lee(1994)Rendell, and Lee]{Rendell:1994kl}
Rendell,~A.~P.; Lee,~T.~J. {Coupled-cluster theory employing approximate
  integrals: An approach to avoid the input/output and storage bottlenecks}.
  \emph{J. Chem. Phys.} \textbf{1994}, \emph{101}, 400--408\relax
\mciteBstWouldAddEndPuncttrue
\mciteSetBstMidEndSepPunct{\mcitedefaultmidpunct}
{\mcitedefaultendpunct}{\mcitedefaultseppunct}\relax
\EndOfBibitem
\bibitem[Bostr{\"o}m \latin{et~al.}(2012)Bostr{\"o}m, Pito{\v{n}}{\'a}k,
  Aquilante, Neogrady, Pedersen, and Lindh]{Bostrom:2012gq}
Bostr{\"o}m,~J.; Pito{\v{n}}{\'a}k,~M.; Aquilante,~F.; Neogrady,~P.;
  Pedersen,~T.~B.; Lindh,~R. {Coupled Cluster and M{\o}ller-Plesset
  Perturbation Theory Calculations of Noncovalent Intermolecular Interactions
  using Density Fitting with Auxiliary Basis Sets from Cholesky
  Decompositions.} \emph{J. Chem. Theory Comput.} \textbf{2012}, \emph{8},
  1921--1928\relax
\mciteBstWouldAddEndPuncttrue
\mciteSetBstMidEndSepPunct{\mcitedefaultmidpunct}
{\mcitedefaultendpunct}{\mcitedefaultseppunct}\relax
\EndOfBibitem
\bibitem[DePrince and Sherrill(2013)DePrince, and Sherrill]{DePrince:2013ki}
DePrince,~A.~E.; Sherrill,~C.~D. {Accuracy and Efficiency of Coupled-Cluster
  Theory Using Density Fitting/Cholesky Decomposition, Frozen Natural Orbitals,
  and a t1-Transformed Hamiltonian}. \emph{J. Chem. Theory Comput.}
  \textbf{2013}, \emph{9}, 2687--2696\relax
\mciteBstWouldAddEndPuncttrue
\mciteSetBstMidEndSepPunct{\mcitedefaultmidpunct}
{\mcitedefaultendpunct}{\mcitedefaultseppunct}\relax
\EndOfBibitem
\bibitem[Epifanovsky \latin{et~al.}(2013)Epifanovsky, Zuev, Feng, Khistyaev,
  Shao, and Krylov]{Epifanovsky:2013gd}
Epifanovsky,~E.; Zuev,~D.; Feng,~X.; Khistyaev,~K.; Shao,~Y.; Krylov,~A.~I.
  {General implementation of the resolution-of-the-identity and Cholesky
  representations of electron repulsion integrals within coupled-cluster and
  equation-of-motion methods: Theory and benchmarks}. \emph{J. Chem. Phys.}
  \textbf{2013}, \emph{139}, 134105\relax
\mciteBstWouldAddEndPuncttrue
\mciteSetBstMidEndSepPunct{\mcitedefaultmidpunct}
{\mcitedefaultendpunct}{\mcitedefaultseppunct}\relax
\EndOfBibitem
\bibitem[Qiu \latin{et~al.}(2017)Qiu, Henderson, Zhao, and
  Scuseria]{Qiu:2017in}
Qiu,~Y.; Henderson,~T.~M.; Zhao,~J.; Scuseria,~G.~E. {Projected coupled cluster
  theory}. \emph{J. Chem. Phys.} \textbf{2017}, \emph{147}, 064111\relax
\mciteBstWouldAddEndPuncttrue
\mciteSetBstMidEndSepPunct{\mcitedefaultmidpunct}
{\mcitedefaultendpunct}{\mcitedefaultseppunct}\relax
\EndOfBibitem
\bibitem[Huntington and Nooijen(2010)Huntington, and
  Nooijen]{Huntington:2010ef}
Huntington,~L. M.~J.; Nooijen,~M. {pCCSD: parameterized coupled-cluster theory
  with single and double excitations.} \emph{J. Chem. Phys.} \textbf{2010},
  \emph{133}, 184109\relax
\mciteBstWouldAddEndPuncttrue
\mciteSetBstMidEndSepPunct{\mcitedefaultmidpunct}
{\mcitedefaultendpunct}{\mcitedefaultseppunct}\relax
\EndOfBibitem
\bibitem[Evangelista and Gauss(2012)Evangelista, and Gauss]{Evangelista:2012hz}
Evangelista,~F.~A.; Gauss,~J. {On the approximation of the
  similarity-transformed Hamiltonian in single-reference and multireference
  coupled cluster theory}. \emph{Chem. Phys.} \textbf{2012}, \emph{401},
  27--35\relax
\mciteBstWouldAddEndPuncttrue
\mciteSetBstMidEndSepPunct{\mcitedefaultmidpunct}
{\mcitedefaultendpunct}{\mcitedefaultseppunct}\relax
\EndOfBibitem
\bibitem[Kats and Manby(2013)Kats, and Manby]{Kats:2013jy}
Kats,~D.; Manby,~F.~R. {Communication: The distinguishable cluster
  approximation}. \emph{J. Chem. Phys.} \textbf{2013}, \emph{139}, 021102\relax
\mciteBstWouldAddEndPuncttrue
\mciteSetBstMidEndSepPunct{\mcitedefaultmidpunct}
{\mcitedefaultendpunct}{\mcitedefaultseppunct}\relax
\EndOfBibitem
\bibitem[Kats(2014)]{Kats:2014bg}
Kats,~D. {Communication: The distinguishable cluster approximation. II. The
  role of orbital relaxation}. \emph{J. Chem. Phys.} \textbf{2014}, \emph{141},
  061101\relax
\mciteBstWouldAddEndPuncttrue
\mciteSetBstMidEndSepPunct{\mcitedefaultmidpunct}
{\mcitedefaultendpunct}{\mcitedefaultseppunct}\relax
\EndOfBibitem
\bibitem[Rishi \latin{et~al.}(2017)Rishi, Perera, Nooijen, and
  Bartlett]{Rishi:2017ky}
Rishi,~V.; Perera,~A.; Nooijen,~M.; Bartlett,~R.~J. {Excited states from
  modified coupled cluster methods: Are they any better than EOM CCSD?}
  \emph{J. Chem. Phys.} \textbf{2017}, \emph{146}, 144104\relax
\mciteBstWouldAddEndPuncttrue
\mciteSetBstMidEndSepPunct{\mcitedefaultmidpunct}
{\mcitedefaultendpunct}{\mcitedefaultseppunct}\relax
\EndOfBibitem
\bibitem[Koch and Kutzelnigg(1981)Koch, and Kutzelnigg]{Koch:1981dx}
Koch,~S.; Kutzelnigg,~W. {Comparison of CEPA and CP-MET methods}. \emph{Theor.
  Chim. Acta} \textbf{1981}, \emph{59}, 387--411\relax
\mciteBstWouldAddEndPuncttrue
\mciteSetBstMidEndSepPunct{\mcitedefaultmidpunct}
{\mcitedefaultendpunct}{\mcitedefaultseppunct}\relax
\EndOfBibitem
\bibitem[Bartlett and Musia{\l}(2006)Bartlett, and Musia{\l}]{Bartlett:2006fh}
Bartlett,~R.~J.; Musia{\l},~M. {Addition by subtraction in coupled-cluster
  theory: A reconsideration of the CC and CI interface and the nCC hierarchy}.
  \emph{J. Chem. Phys.} \textbf{2006}, \emph{125}, 204105\relax
\mciteBstWouldAddEndPuncttrue
\mciteSetBstMidEndSepPunct{\mcitedefaultmidpunct}
{\mcitedefaultendpunct}{\mcitedefaultseppunct}\relax
\EndOfBibitem
\bibitem[Roos \latin{et~al.}(1980)Roos, Taylor, and Siegbahn]{Roos:1980fd}
Roos,~B.~O.; Taylor,~P.~R.; Siegbahn,~P. E.~M. {A complete active space SCF
  method (CASSCF) using a density matrix formulated super-CI approach}.
  \emph{Chem. Phys.} \textbf{1980}, \emph{48}, 157--173\relax
\mciteBstWouldAddEndPuncttrue
\mciteSetBstMidEndSepPunct{\mcitedefaultmidpunct}
{\mcitedefaultendpunct}{\mcitedefaultseppunct}\relax
\EndOfBibitem
\bibitem[Mukherjee(1997)]{Mukherjee:1997kf}
Mukherjee,~D. {Normal ordering and a Wick-like reduction theorem for fermions
  with respect to a multi-determinantal reference state}. \emph{Chem. Phys.
  Lett.} \textbf{1997}, \emph{274}, 561--566\relax
\mciteBstWouldAddEndPuncttrue
\mciteSetBstMidEndSepPunct{\mcitedefaultmidpunct}
{\mcitedefaultendpunct}{\mcitedefaultseppunct}\relax
\EndOfBibitem
\bibitem[Kutzelnigg and Mukherjee(1997)Kutzelnigg, and
  Mukherjee]{Kutzelnigg:1997dp}
Kutzelnigg,~W.; Mukherjee,~D. {Normal order and extended Wick theorem for a
  multiconfiguration reference wave function}. \emph{J. Chem. Phys.}
  \textbf{1997}, \emph{107}, 432--449\relax
\mciteBstWouldAddEndPuncttrue
\mciteSetBstMidEndSepPunct{\mcitedefaultmidpunct}
{\mcitedefaultendpunct}{\mcitedefaultseppunct}\relax
\EndOfBibitem
\bibitem[Kutzelnigg \latin{et~al.}(2010)Kutzelnigg, Shamasundar, and
  Mukherjee]{Kutzelnigg:2010iu}
Kutzelnigg,~W.; Shamasundar,~K.~R.; Mukherjee,~D. {Spinfree formulation of
  reduced density matrices, density cumulants and generalised normal ordering}.
  \emph{Mol. Phys.} \textbf{2010}, \emph{108}, 433--451\relax
\mciteBstWouldAddEndPuncttrue
\mciteSetBstMidEndSepPunct{\mcitedefaultmidpunct}
{\mcitedefaultendpunct}{\mcitedefaultseppunct}\relax
\EndOfBibitem
\bibitem[Kong \latin{et~al.}(2010)Kong, Nooijen, and Mukherjee]{Kong:2010kg}
Kong,~L.; Nooijen,~M.; Mukherjee,~D. {An algebraic proof of generalized Wick
  theorem}. \emph{J. Chem. Phys.} \textbf{2010}, \emph{132}, 234107\relax
\mciteBstWouldAddEndPuncttrue
\mciteSetBstMidEndSepPunct{\mcitedefaultmidpunct}
{\mcitedefaultendpunct}{\mcitedefaultseppunct}\relax
\EndOfBibitem
\bibitem[Sinha \latin{et~al.}(2013)Sinha, Maitra, and Mukherjee]{Sinha:2013dx}
Sinha,~D.; Maitra,~R.; Mukherjee,~D. {Generalized antisymmetric ordered
  products, generalized normal ordered products, ordered and ordinary cumulants
  and their use in many electron correlation problem}. \emph{Comput. Theor.
  Chem.} \textbf{2013}, \emph{1003}, 62--70\relax
\mciteBstWouldAddEndPuncttrue
\mciteSetBstMidEndSepPunct{\mcitedefaultmidpunct}
{\mcitedefaultendpunct}{\mcitedefaultseppunct}\relax
\EndOfBibitem
\bibitem[FOR(2019)]{FORTE2019}
Forte, a suite of quantum chemistry methods for strongly correlated electrons.
  For current version see \url{https://github.com/evangelistalab/forte},
  2019\relax
\mciteBstWouldAddEndPuncttrue
\mciteSetBstMidEndSepPunct{\mcitedefaultmidpunct}
{\mcitedefaultendpunct}{\mcitedefaultseppunct}\relax
\EndOfBibitem
\bibitem[AMB(2018)]{AMBIT2018}
Ambit, a C++ library for the implementation of tensor product calculations
  through a clean, concise user interface. For current version see
  \url{https://github.com/jturney/ambit}, 2018\relax
\mciteBstWouldAddEndPuncttrue
\mciteSetBstMidEndSepPunct{\mcitedefaultmidpunct}
{\mcitedefaultendpunct}{\mcitedefaultseppunct}\relax
\EndOfBibitem
\bibitem[Parrish \latin{et~al.}(2017)Parrish, Burns, Smith, Simmonett,
  DePrince, Hohenstein, Bozkaya, Sokolov, Di~Remigio, Richard, Gonthier, James,
  McAlexander, Kumar, Saitow, Wang, Pritchard, Verma, Schaefer, Patkowski,
  King, Valeev, Evangelista, Turney, Crawford, and Sherrill]{Parrish:2017hg}
Parrish,~R.~M.; Burns,~L.~A.; Smith,~D. G.~A.; Simmonett,~A.~C.;
  DePrince,~A.~E.; Hohenstein,~E.~G.; Bozkaya,~U.; Sokolov,~A.~Y.;
  Di~Remigio,~R.; Richard,~R.~M.; Gonthier,~J.~F.; James,~A.~M.;
  McAlexander,~H.~R.; Kumar,~A.; Saitow,~M.; Wang,~X.; Pritchard,~B.~P.;
  Verma,~P.; Schaefer,~H.~F.; Patkowski,~K.; King,~R.~A.; Valeev,~E.~F.;
  Evangelista,~F.~A.; Turney,~J.~M.; Crawford,~T.~D.; Sherrill,~C.~D. {Psi4
  1.1: An Open-Source Electronic Structure Program Emphasizing Automation,
  Advanced Libraries, and Interoperability}. \emph{J. Chem. Theory Comput.}
  \textbf{2017}, \emph{13}, 3185--3197\relax
\mciteBstWouldAddEndPuncttrue
\mciteSetBstMidEndSepPunct{\mcitedefaultmidpunct}
{\mcitedefaultendpunct}{\mcitedefaultseppunct}\relax
\EndOfBibitem
\bibitem[Smith \latin{et~al.}(2018)Smith, Burns, Sirianni, Nascimento, Kumar,
  James, Schriber, Zhang, Zhang, Abbott, Berquist, Lechner, Cunha, Heide,
  Waldrop, Takeshita, Alenaizan, Neuhauser, King, Simmonett, Turney, Schaefer,
  Evangelista, DePrince, Crawford, Patkowski, and Sherrill]{Smith:2018ip}
Smith,~D. G.~A.; Burns,~L.~A.; Sirianni,~D.~A.; Nascimento,~D.~R.; Kumar,~A.;
  James,~A.~M.; Schriber,~J.~B.; Zhang,~T.; Zhang,~B.; Abbott,~A.~S.;
  Berquist,~E.~J.; Lechner,~M.~H.; Cunha,~L.~A.; Heide,~A.~G.; Waldrop,~J.~M.;
  Takeshita,~T.~Y.; Alenaizan,~A.; Neuhauser,~D.; King,~R.~A.;
  Simmonett,~A.~C.; Turney,~J.~M.; Schaefer,~H.~F.; Evangelista,~F.~A.;
  DePrince,~A.~E.; Crawford,~T.~D.; Patkowski,~K.; Sherrill,~C.~D. {Psi4NumPy:
  An Interactive Quantum Chemistry Programming Environment for Reference
  Implementations and Rapid Development}. \emph{J. Chem. Theory Comput.}
  \textbf{2018}, \emph{14}, 3504--3511\relax
\mciteBstWouldAddEndPuncttrue
\mciteSetBstMidEndSepPunct{\mcitedefaultmidpunct}
{\mcitedefaultendpunct}{\mcitedefaultseppunct}\relax
\EndOfBibitem
\bibitem[Hohenstein \latin{et~al.}(2012)Hohenstein, Parrish, and
  Mart{\'\i}nez]{Hohenstein:2012hk}
Hohenstein,~E.~G.; Parrish,~R.~M.; Mart{\'\i}nez,~T.~J. {Tensor
  hypercontraction density fitting. I. Quartic scaling second- and third-order
  M{\o}ller-Plesset perturbation theory.} \emph{J. Chem. Phys.} \textbf{2012},
  \emph{137}, 044103\relax
\mciteBstWouldAddEndPuncttrue
\mciteSetBstMidEndSepPunct{\mcitedefaultmidpunct}
{\mcitedefaultendpunct}{\mcitedefaultseppunct}\relax
\EndOfBibitem
\bibitem[Huber and Herzberg(1979)Huber, and Herzberg]{Huber:1979cc}
Huber,~K.~P.; Herzberg,~G. \emph{Molecular Spectra and Molecular Structure};
  Springer: Boston, MA, 1979\relax
\mciteBstWouldAddEndPuncttrue
\mciteSetBstMidEndSepPunct{\mcitedefaultmidpunct}
{\mcitedefaultendpunct}{\mcitedefaultseppunct}\relax
\EndOfBibitem
\bibitem[Irikura(2007)]{Irikura:2007jg}
Irikura,~K.~K. {Experimental Vibrational Zero-Point Energies: Diatomic
  Molecules}. \emph{J. Phys. Chem. Ref. Data} \textbf{2007}, \emph{36},
  389--397\relax
\mciteBstWouldAddEndPuncttrue
\mciteSetBstMidEndSepPunct{\mcitedefaultmidpunct}
{\mcitedefaultendpunct}{\mcitedefaultseppunct}\relax
\EndOfBibitem
\bibitem[Dunning~Jr(1989)]{Dunning:1989bx}
Dunning~Jr,~T.~H. {Gaussian basis sets for use in correlated molecular
  calculations. I. The atoms boron through neon and hydrogen}. \emph{J. Chem.
  Phys.} \textbf{1989}, \emph{90}, 1007--1018\relax
\mciteBstWouldAddEndPuncttrue
\mciteSetBstMidEndSepPunct{\mcitedefaultmidpunct}
{\mcitedefaultendpunct}{\mcitedefaultseppunct}\relax
\EndOfBibitem
\bibitem[Woon and Dunning(1994)Woon, and Dunning]{Woon:1994jq}
Woon,~D.~E.; Dunning,~T.~H. {Gaussian basis sets for use in correlated
  molecular calculations. IV. Calculation of static electrical response
  properties}. \emph{J. Chem. Phys.} \textbf{1994}, \emph{100}, 2975\relax
\mciteBstWouldAddEndPuncttrue
\mciteSetBstMidEndSepPunct{\mcitedefaultmidpunct}
{\mcitedefaultendpunct}{\mcitedefaultseppunct}\relax
\EndOfBibitem
\bibitem[Weigend and Ahlrichs(2005)Weigend, and Ahlrichs]{Weigend:2005gx}
Weigend,~F.; Ahlrichs,~R. {Balanced basis sets of split valence, triple zeta
  valence and quadruple zeta valence quality for H to Rn: Design and assessment
  of accuracy}. \emph{Phys. Chem. Chem. Phys.} \textbf{2005}, \emph{7},
  3297--3305\relax
\mciteBstWouldAddEndPuncttrue
\mciteSetBstMidEndSepPunct{\mcitedefaultmidpunct}
{\mcitedefaultendpunct}{\mcitedefaultseppunct}\relax
\EndOfBibitem
\bibitem[Weigend(2008)]{Weigend:2008df}
Weigend,~F. {Hartree{\textendash}Fock exchange fitting basis sets for H to Rn}.
  \emph{J. Comput. Chem.} \textbf{2008}, \emph{29}, 167--175\relax
\mciteBstWouldAddEndPuncttrue
\mciteSetBstMidEndSepPunct{\mcitedefaultmidpunct}
{\mcitedefaultendpunct}{\mcitedefaultseppunct}\relax
\EndOfBibitem
\bibitem[Weigend \latin{et~al.}(2002)Weigend, K{\"o}hn, and
  H{\"a}ttig]{Weigend:2002jp}
Weigend,~F.; K{\"o}hn,~A.; H{\"a}ttig,~C. {Efficient use of the correlation
  consistent basis sets in resolution of the identity MP2 calculations}.
  \emph{J. Chem. Phys.} \textbf{2002}, \emph{116}, 3175--3183\relax
\mciteBstWouldAddEndPuncttrue
\mciteSetBstMidEndSepPunct{\mcitedefaultmidpunct}
{\mcitedefaultendpunct}{\mcitedefaultseppunct}\relax
\EndOfBibitem
\bibitem[H{\"a}ttig(2005)]{Hattig:2005dm}
H{\"a}ttig,~C. {Optimization of auxiliary basis sets for RI-MP2 and RI-CC2
  calculations: Core{\textendash}valence and quintuple-$\zeta$ basis sets for H
  to Ar and QZVPP basis sets for Li to Kr}. \emph{Phys. Chem. Chem. Phys.}
  \textbf{2005}, \emph{7}, 59--66\relax
\mciteBstWouldAddEndPuncttrue
\mciteSetBstMidEndSepPunct{\mcitedefaultmidpunct}
{\mcitedefaultendpunct}{\mcitedefaultseppunct}\relax
\EndOfBibitem
\bibitem[Shen and Piecuch(2012)Shen, and Piecuch]{Shen:2012kn}
Shen,~J.; Piecuch,~P. {Combining active-space coupled-cluster methods with
  moment energy corrections via the CC(P;Q) methodology, with benchmark
  calculations for biradical transition states}. \emph{J. Chem. Phys.}
  \textbf{2012}, \emph{136}, 144104\relax
\mciteBstWouldAddEndPuncttrue
\mciteSetBstMidEndSepPunct{\mcitedefaultmidpunct}
{\mcitedefaultendpunct}{\mcitedefaultseppunct}\relax
\EndOfBibitem
\bibitem[Whitman and Carpenter(1982)Whitman, and Carpenter]{Whitman:1982hy}
Whitman,~D.~W.; Carpenter,~B.~K. {Limits on the activation parameters for
  automerization of cyclobutadiene-1,2-d$_{2}$}. \emph{J. Am. Chem. Soc.}
  \textbf{1982}, \emph{104}, 6473--6474\relax
\mciteBstWouldAddEndPuncttrue
\mciteSetBstMidEndSepPunct{\mcitedefaultmidpunct}
{\mcitedefaultendpunct}{\mcitedefaultseppunct}\relax
\EndOfBibitem
\bibitem[Nakamura \latin{et~al.}(1989)Nakamura, Osamura, and
  Iwata]{Nakamura:1989ff}
Nakamura,~K.; Osamura,~Y.; Iwata,~S. {Second-order jahn-teller effect of
  cyclobutadiene in low-lying states. An MCSCF study}. \emph{Chem. Phys.}
  \textbf{1989}, \emph{136}, 67--77\relax
\mciteBstWouldAddEndPuncttrue
\mciteSetBstMidEndSepPunct{\mcitedefaultmidpunct}
{\mcitedefaultendpunct}{\mcitedefaultseppunct}\relax
\EndOfBibitem
\bibitem[Balkov{\'a} and Bartlett(1998)Balkov{\'a}, and
  Bartlett]{Balkova:1994kp}
Balkov{\'a},~A.; Bartlett,~R.~J. {A multireference coupled-cluster study of the
  ground state and lowest excited states of cyclobutadiene}. \emph{J. Chem.
  Phys.} \textbf{1998}, \emph{101}, 8972--8987\relax
\mciteBstWouldAddEndPuncttrue
\mciteSetBstMidEndSepPunct{\mcitedefaultmidpunct}
{\mcitedefaultendpunct}{\mcitedefaultseppunct}\relax
\EndOfBibitem
\bibitem[Sancho-Garc{\'\i}a \latin{et~al.}(2000)Sancho-Garc{\'\i}a, Pittner,
  {\v C}{\'a}rsky, and Huba{\v c}]{SanchoGarcia:2000dx}
Sancho-Garc{\'\i}a,~J.~C.; Pittner,~J.; {\v C}{\'a}rsky,~P.; Huba{\v c},~I.
  {Multireference coupled-cluster calculations on the energy of activation in
  the automerization of cyclobutadiene: Assessment of the state-specific
  multireference Brillouin{\textendash}Wigner theory}. \emph{J. Chem. Phys.}
  \textbf{2000}, \emph{112}, 8785--8788\relax
\mciteBstWouldAddEndPuncttrue
\mciteSetBstMidEndSepPunct{\mcitedefaultmidpunct}
{\mcitedefaultendpunct}{\mcitedefaultseppunct}\relax
\EndOfBibitem
\bibitem[Levchenko and Krylov(2003)Levchenko, and Krylov]{Levchenko:2003hq}
Levchenko,~S.~V.; Krylov,~A.~I. {Equation-of-motion spin-flip coupled-cluster
  model with single and double substitutions: Theory and application to
  cyclobutadiene}. \emph{J. Chem. Phys.} \textbf{2003}, \emph{120},
  175--185\relax
\mciteBstWouldAddEndPuncttrue
\mciteSetBstMidEndSepPunct{\mcitedefaultmidpunct}
{\mcitedefaultendpunct}{\mcitedefaultseppunct}\relax
\EndOfBibitem
\bibitem[Eckert-Maksi{\'c} \latin{et~al.}(2006)Eckert-Maksi{\'c}, Vazdar,
  Barbatti, Lischka, and Maksi{\'c}]{EckertMaksic:2006ex}
Eckert-Maksi{\'c},~M.; Vazdar,~M.; Barbatti,~M.; Lischka,~H.; Maksi{\'c},~Z.~B.
  {Automerization reaction of cyclobutadiene and its barrier height: An ab
  initio benchmark multireference average-quadratic coupled cluster study}.
  \emph{J. Chem. Phys.} \textbf{2006}, \emph{125}, 064310\relax
\mciteBstWouldAddEndPuncttrue
\mciteSetBstMidEndSepPunct{\mcitedefaultmidpunct}
{\mcitedefaultendpunct}{\mcitedefaultseppunct}\relax
\EndOfBibitem
\bibitem[Lyakh \latin{et~al.}(2011)Lyakh, Lotrich, and Bartlett]{Lyakh:2011iq}
Lyakh,~D.~I.; Lotrich,~V.~F.; Bartlett,~R.~J. {The
  {\textquoteleft}tailored{\textquoteright} CCSD(T) description of the
  automerization of cyclobutadiene}. \emph{Chem. Phys. Lett.} \textbf{2011},
  \emph{501}, 166--171\relax
\mciteBstWouldAddEndPuncttrue
\mciteSetBstMidEndSepPunct{\mcitedefaultmidpunct}
{\mcitedefaultendpunct}{\mcitedefaultseppunct}\relax
\EndOfBibitem
\bibitem[Shen \latin{et~al.}(2008)Shen, Fang, Li, and Jiang]{Shen:2008kv}
Shen,~J.; Fang,~T.; Li,~S.; Jiang,~Y. {Performance of Block Correlated Coupled
  Cluster Method with the CASSCF Reference Function for the Prediction of
  Activation Barriers, Spectroscopic Constants in Diatomic Molecules, and
  Singlet{\textminus}Triplet Gaps in Diradicals}. \emph{J. Phys. Chem. A}
  \textbf{2008}, \emph{112}, 12518--12525\relax
\mciteBstWouldAddEndPuncttrue
\mciteSetBstMidEndSepPunct{\mcitedefaultmidpunct}
{\mcitedefaultendpunct}{\mcitedefaultseppunct}\relax
\EndOfBibitem
\bibitem[Li and Paldus(2009)Li, and Paldus]{Li:2009eq}
Li,~X.; Paldus,~J. {Accounting for the exact degeneracy and quasidegeneracy in
  the automerization of cyclobutadiene via multireference coupled-cluster
  methods}. \emph{J. Chem. Phys.} \textbf{2009}, \emph{131}, 114103\relax
\mciteBstWouldAddEndPuncttrue
\mciteSetBstMidEndSepPunct{\mcitedefaultmidpunct}
{\mcitedefaultendpunct}{\mcitedefaultseppunct}\relax
\EndOfBibitem
\bibitem[Bhaskaran-Nair \latin{et~al.}(2008)Bhaskaran-Nair, Demel, and
  Pittner]{BhaskaranNair:2008dj}
Bhaskaran-Nair,~K.; Demel,~O.; Pittner,~J. {Multireference state-specific
  Mukherjee{\textquoteright}s coupled cluster method with noniterative
  triexcitations}. \emph{J. Chem. Phys.} \textbf{2008}, \emph{129},
  184105\relax
\mciteBstWouldAddEndPuncttrue
\mciteSetBstMidEndSepPunct{\mcitedefaultmidpunct}
{\mcitedefaultendpunct}{\mcitedefaultseppunct}\relax
\EndOfBibitem
\bibitem[Demel \latin{et~al.}(2008)Demel, Shamasundar, Kong, and
  Nooijen]{Demel:2008ey}
Demel,~O.; Shamasundar,~K.~R.; Kong,~L.; Nooijen,~M. {Application of Double
  Ionization State-Specific Equation of Motion Coupled Cluster Method to
  Organic Diradicals}. \emph{J. Phys. Chem. A} \textbf{2008}, \emph{112},
  11895--11902\relax
\mciteBstWouldAddEndPuncttrue
\mciteSetBstMidEndSepPunct{\mcitedefaultmidpunct}
{\mcitedefaultendpunct}{\mcitedefaultseppunct}\relax
\EndOfBibitem
\bibitem[Wu \latin{et~al.}(2012)Wu, Mo, Evangelista, and Schleyer]{Wu:2012ek}
Wu,~J. I.-C.; Mo,~Y.; Evangelista,~F.~A.; Schleyer,~P. v.~R. {Is cyclobutadiene
  really highly destabilized by antiaromaticity?} \emph{Chem. Commun.}
  \textbf{2012}, \emph{48}, 8437--8439\relax
\mciteBstWouldAddEndPuncttrue
\mciteSetBstMidEndSepPunct{\mcitedefaultmidpunct}
{\mcitedefaultendpunct}{\mcitedefaultseppunct}\relax
\EndOfBibitem
\bibitem[Mahapatra \latin{et~al.}(1999)Mahapatra, Datta, and
  Mukherjee]{Mahapatra:1999bp}
Mahapatra,~U.~S.; Datta,~B.; Mukherjee,~D. {Molecular Applications of a
  Size-Consistent State-Specific Multireference Perturbation Theory with
  Relaxed Model-Space Coefficients}. \emph{J. Phys. Chem. A} \textbf{1999},
  \emph{103}, 1822--1830\relax
\mciteBstWouldAddEndPuncttrue
\mciteSetBstMidEndSepPunct{\mcitedefaultmidpunct}
{\mcitedefaultendpunct}{\mcitedefaultseppunct}\relax
\EndOfBibitem
\bibitem[Chattopadhyay \latin{et~al.}(2000)Chattopadhyay, Mahapatra, and
  Mukherjee]{Chattopadhyay:2000io}
Chattopadhyay,~S.; Mahapatra,~U.~S.; Mukherjee,~D. {Development of a linear
  response theory based on a state-specific multireference coupled cluster
  formalism}. \emph{J. Chem. Phys.} \textbf{2000}, \emph{112}, 7939--7952\relax
\mciteBstWouldAddEndPuncttrue
\mciteSetBstMidEndSepPunct{\mcitedefaultmidpunct}
{\mcitedefaultendpunct}{\mcitedefaultseppunct}\relax
\EndOfBibitem
\bibitem[Chattopadhyay \latin{et~al.}(2004)Chattopadhyay, Pahari, Mukherjee,
  and Mahapatra]{Chattopadhyay:2004fw}
Chattopadhyay,~S.; Pahari,~D.; Mukherjee,~D.; Mahapatra,~U.~S. {A
  state-specific approach to multireference coupled electron-pair approximation
  like methods: Development and applications}. \emph{J. Chem. Phys.}
  \textbf{2004}, \emph{120}, 5968--5986\relax
\mciteBstWouldAddEndPuncttrue
\mciteSetBstMidEndSepPunct{\mcitedefaultmidpunct}
{\mcitedefaultendpunct}{\mcitedefaultseppunct}\relax
\EndOfBibitem
\bibitem[Pahari \latin{et~al.}(2004)Pahari, Chattopadhyay, Deb, and
  Mukherjee]{Pahari:2004ih}
Pahari,~D.; Chattopadhyay,~S.; Deb,~A.; Mukherjee,~D. {An orbital-invariant
  coupled electron-pair like approximant to a state-specific multi-reference
  coupled cluster formalism}. \emph{Chem. Phys. Lett.} \textbf{2004},
  \emph{386}, 307--312\relax
\mciteBstWouldAddEndPuncttrue
\mciteSetBstMidEndSepPunct{\mcitedefaultmidpunct}
{\mcitedefaultendpunct}{\mcitedefaultseppunct}\relax
\EndOfBibitem
\bibitem[Evangelista \latin{et~al.}(2006)Evangelista, Allen, and
  Schaefer]{Evangelista:2006gf}
Evangelista,~F.~A.; Allen,~W.~D.; Schaefer,~H.~F. {High-order excitations in
  state-universal and state-specific multireference coupled cluster theories:
  Model systems}. \emph{J. Chem. Phys.} \textbf{2006}, \emph{125}, 154113\relax
\mciteBstWouldAddEndPuncttrue
\mciteSetBstMidEndSepPunct{\mcitedefaultmidpunct}
{\mcitedefaultendpunct}{\mcitedefaultseppunct}\relax
\EndOfBibitem
\bibitem[Evangelista \latin{et~al.}(2008)Evangelista, Simmonett, Allen,
  Schaefer, and Gauss]{Evangelista:2008gv}
Evangelista,~F.~A.; Simmonett,~A.~C.; Allen,~W.~D.; Schaefer,~H.~F.; Gauss,~J.
  {Triple excitations in state-specific multireference coupled cluster theory:
  Application of Mk-MRCCSDT and Mk-MRCCSDT-n methods to model systems}.
  \emph{J. Chem. Phys.} \textbf{2008}, \emph{128}, 124104\relax
\mciteBstWouldAddEndPuncttrue
\mciteSetBstMidEndSepPunct{\mcitedefaultmidpunct}
{\mcitedefaultendpunct}{\mcitedefaultseppunct}\relax
\EndOfBibitem
\bibitem[Evangelista \latin{et~al.}(2010)Evangelista, Prochnow, Gauss, and
  Schaefer]{Evangelista:2010cq}
Evangelista,~F.~A.; Prochnow,~E.; Gauss,~J.; Schaefer,~H.~F. {Perturbative
  triples corrections in state-specific multireference coupled cluster theory.}
  \emph{J. Chem. Phys.} \textbf{2010}, \emph{132}, 074107\relax
\mciteBstWouldAddEndPuncttrue
\mciteSetBstMidEndSepPunct{\mcitedefaultmidpunct}
{\mcitedefaultendpunct}{\mcitedefaultseppunct}\relax
\EndOfBibitem
\bibitem[Taube and Bartlett(2009)Taube, and Bartlett]{Taube:2009jz}
Taube,~A.~G.; Bartlett,~R.~J. {Rethinking linearized coupled-cluster theory}.
  \emph{J. Chem. Phys.} \textbf{2009}, \emph{130}, 144112\relax
\mciteBstWouldAddEndPuncttrue
\mciteSetBstMidEndSepPunct{\mcitedefaultmidpunct}
{\mcitedefaultendpunct}{\mcitedefaultseppunct}\relax
\EndOfBibitem
\bibitem[Chiles and Dykstra(1981)Chiles, and Dykstra]{Chiles:1981jg}
Chiles,~R.~A.; Dykstra,~C.~E. {An electron pair operator approach to coupled
  cluster wave functions. Application to He$_{2}$, Be$_{2}$, and Mg$_{2}$ and
  comparison with CEPA methods}. \emph{J. Chem. Phys.} \textbf{1981},
  \emph{74}, 4544--4556\relax
\mciteBstWouldAddEndPuncttrue
\mciteSetBstMidEndSepPunct{\mcitedefaultmidpunct}
{\mcitedefaultendpunct}{\mcitedefaultseppunct}\relax
\EndOfBibitem
\bibitem[Handy \latin{et~al.}(1989)Handy, Pople, Head-Gordon, Raghavachari, and
  Trucks]{Handy:1989fn}
Handy,~N.~C.; Pople,~J.~A.; Head-Gordon,~M.; Raghavachari,~K.; Trucks,~G.~W.
  {Size-consistent Brueckner theory limited to double substitutions}.
  \emph{Chem. Phys. Lett.} \textbf{1989}, \emph{164}, 185--192\relax
\mciteBstWouldAddEndPuncttrue
\mciteSetBstMidEndSepPunct{\mcitedefaultmidpunct}
{\mcitedefaultendpunct}{\mcitedefaultseppunct}\relax
\EndOfBibitem
\bibitem[Stanton \latin{et~al.}(1998)Stanton, Gauss, and
  Bartlett]{Stanton:1998fk}
Stanton,~J.~F.; Gauss,~J.; Bartlett,~R.~J. {On the choice of orbitals for
  symmetry breaking problems with application to NO$_{3}$}. \emph{J. Chem.
  Phys.} \textbf{1998}, \emph{97}, 5554--5559\relax
\mciteBstWouldAddEndPuncttrue
\mciteSetBstMidEndSepPunct{\mcitedefaultmidpunct}
{\mcitedefaultendpunct}{\mcitedefaultseppunct}\relax
\EndOfBibitem
\end{mcitethebibliography}

\end{document}